\documentclass[pdftex,onecolumn,epjc3]{svjour3}          

\RequirePackage[T1]{fontenc}

\smartqed  
\usepackage{subcaption} 
\usepackage{microtype}
\usepackage{amsmath}
\usepackage{amssymb}
\usepackage{xcolor}
\usepackage{float} 

\RequirePackage{graphicx}
\RequirePackage{mathptmx}      
\RequirePackage{flushend}
\RequirePackage[numbers,sort&compress]{natbib}
\RequirePackage[colorlinks,citecolor=blue,urlcolor=blue,linkcolor=blue]{hyperref}

\journalname{Eur. Phys. J. C}

\begin{document}
\title{Estimating Cosmological Parameters and Reconstructing Hubble Constant with Artificial Neural Networks: A Test with covariance matrix and mock H(z)}

\author{Jie-feng Chen\thanksref{addr1,addr2,addr3}
        \and
        Tong-Jie Zhang\thanksref{t1, e1,addr2,addr3} 
		\and
		Peng He\thanksref{addr4}
		\and		
		Tingting Zhang\thanksref{addr5}
		and
		Jie Zhang\thanksref{e2,addr1}
}

\thankstext[$\star$]{t1}{Corresponding Author}
\thankstext{e1}{e-mail: tjzhang@bnu.edu.cn}
\thankstext{e2}{e-mail: zhangjie@qlnu.edu.cn}

\institute{School of arts and sciences, Shanghai Dianji University, Shanghai, 200240, China\label{addr1}
          \and Institute for Frontiers in Astronomy and Astrophysics, Beijing Normal University, Beijing 102206,						  China\label{addr2}
          \and School of Physics and Astronomy, Beijing Normal University, Beijing 100875, China\label{addr3}
		  \and Bureau of Frontier Sciences and Basic Research, Chinese Academy of Sciences, Beijing 100190, China\label{addr4}
		  \and College of Command and Control Engineering, PLA Army Engineering University, Nanjing 210000, China\label{addr5}
}

\date{Received: date / Accepted: date}

\maketitle

\begin{abstract}
In this work, we reconstruct the $H(z)$ based on observational Hubble data with Artificial Neural Network, then estimate the cosmological parameters and the Hubble constant. The training data we used are covariance matrix and mock $H(z)$, which are generated based on the real OHD data and Gaussian Process(GP). The use of the covariance matrix propagates the correlated uncertainties and improves training efficiency. Using the reconstructed H(z) data, we first determine the Hubble constant and compare it with CMB-based measurements. To constrain cosmological parameters, we sample on the reconstructed data and calculate the corresponding posterior distributions with Markov Chain Monte Carlo (MCMC). Through comprehensive statistical comparisons, we demonstrate that the parameter estimation using reconstructed samples achieves comparable statistical accuracy to the result derived from real OHD data.
\end{abstract}

\section{Introduction}
\label{sec:intro}

The accelerating expansion of the universe is one of the most important discoveries in modern cosmology research. In order to explain this phenomenon, constraining cosmological parameters and estimating the Hubble constant have long been fundamental tasks in cosmology.

To achieve these goals, researchers rely on a variety of observational datasets that probe different aspects of the universe. For instance, current datasets include Damped Lyman - $\alpha$  Absorber (DLA) of HI 21 cm system \citep{lu2022toward, lu2022statistical}, observational Hubble parameter data (OHD, \citet{2017Bayesian}), type Ia supernovae (SNe Ia, \cite{2017The}), cosmic microwave background (\cite{2017The}), and large-scale structures \citep{2020SCPMA..6310412P}. Bseides constraining the cosmological parameters \citep{niu2023cosmological, zhang2023kernel}, with the observational data, we can understand the neutrino condensation \citep{yu2017differential}, searching for extraterrestrial intelligence \citep{tao2022sensitive,luan2023multibeam} and so forth. However, in some occasions, because of the insufficiency of the existing data, we need to reconstruct or interpolate the observational data, according to the real observational datas and hypotheses.

At the same time, in the past few decades, the artificial neural networks (ANN) developed rapidly \citep{jimenez2003constraints}. The introduction of deep learning architectures enabled training of very deep neural networks with numerous layers. Deep learning models like Transformer networks, GANs (Generative Adversarial Networks, \cite{goodfellow2014generative}), and BERT (Bidirectional Encoder Representations from Transformers \citep{kenton2019bert}) have demonstrated remarkable performance in different applications. Therefore, more and more astrophysics tasks started to apply ANNs, such as processing large scale astronomical data \citep{cabayol2020pau}, constraining the cosmological parameters \citep{2021Likelihood,chen2023test} and data reconstruction \citep{zhang2023non}.

Reconstruction methods are essential in cosmology, where observational data are often sparse and unevenly distributed. By reconstructing quantities such as the Hubble parameter, we are able to obtain a continuous representation of the Universe’s evolution, enabling more reliable inference of cosmological parameters. Neural networks have recently been explored as powerful non-parametric tools for cosmological reconstructions, particularly in handling sparse or noisy data. As highlighted in the Cosmoverse White Paper\cite{2025arXiv250401669D}, they offer a flexible alternative to traditional methods, and have been applied to reconstruct cosmological functions. In this work, we attempt to reconstruct the Hubble parameter in the redshift range of $z\in [0,2]$, and then estimate the cosmological parameters and reconstruct the Hubble constant with the reconstructed data. We proposed a new method to complete reconstruction in our work, combining the mock data and the covariance matrix. In particular, \citep{Zhang_2024}, \citep{wang2020reconstructing}, and \citep{G_mez_Vargas_2023} have previously reconstructed the same H(z) data. However, covariance matrix was not considered in the reconstruction process in \citep{wang2020reconstructing}. Though \citep{G_mez_Vargas_2023} and \citep{Zhang_2024} took the covariance matrix into account, they just generated covariance matrices based on the existing OHD data points. In our work, the covariance matrix we constructed yields non-zero entries in regions lacking observational data. One of the advantage of our method is the use of the covariance matrix ensures that correlated uncertainties are properly propagated, which could improve the robustness of parameter estimation and reduce training time. Besides, the framework (such as the generation of the covariance matrix) is readily extendable and expected to become more powerful when larger and more precise OHD data points (e.g., CSST, Euclid) become available.

This paper is organized as follows: In section \ref{sec:2}, firstly we briefly introduce the basic knowledge of the artificial neural network, and how we realize the reconstruction with ANN. In section \ref{sec:3}, we introduce the OHD data we use in this project, and how we generate the mock OHD data. Besides, we also show the reason why we added an additional covariance matrix to the neural network, and how we built it. In section \ref{sec:4}, we show the reconstructed Hubble data generated by the neural network, and we calculate the Hubble constant \citep{jackson2015hubble,freedman2010hubble} with the reconstructed data. Meanwhile, we constrain the cosmological parameters and do quantitative analysis (Kullback-Leibler divergence and Figure of Merit). Finally in Section \ref{sec:5}, we discuss and conclude.

\section{Methodology}
\label{sec:2}

We first review basic principles of ANNs and CNNs in this section, then introduce the specific neural network architecture we employed in our work.

\subsection{Neural Network}

Artificial Neural Networks (ANNs) \citep{yegnanarayana2009artificial} are computational models inspired by biological neural systems. They are typically used to approximate complex functions (mappings) between inputs and outputs through a set of interconnected layers.

\subsubsection{Network Structure and Activation Functions}\label{Layers}

A basic ANN typically consists of an input layer, one or more hidden layers, and an output layer. The input layer receives input data, which is then transformed through the hidden layers using weighted connections and nonlinear activation functions, ultimately the output layer produces the prediction (output).

\begin{figure}[h]
    \centering
    \begin{subfigure}{0.25\textwidth} 
        \centering
        \includegraphics[width=\textwidth]{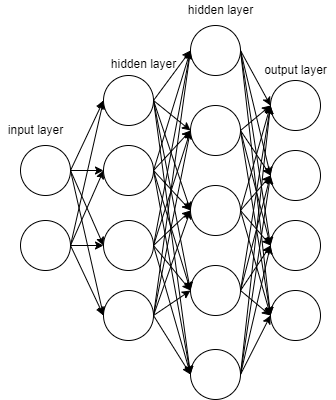} 
        \caption{Basic structure of an artificial neural network (ANN), including input layer, hidden layer(s), and output layer. Practical ANNs typically have more complex architectures.}
        \label{fig:ANN}
    \end{subfigure}
    \hfill 
    \begin{subfigure}{0.25\textwidth}
        \centering
        \includegraphics[width=\textwidth]{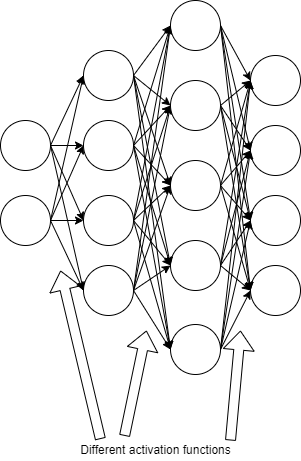}
        \caption{As shown above, activation functions are applied to the connections between units in adjacent layers.}
        \label{fig:activition}
    \end{subfigure}
    \hfill 
    \begin{subfigure}{0.28\textwidth}
        \centering
        \includegraphics[width=\textwidth]{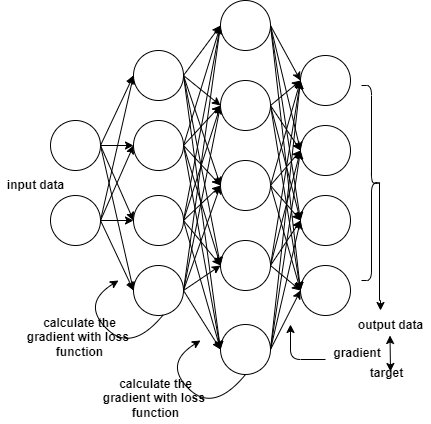}
        \caption{During backpropagation, the algorithm first computes the loss between the network's output and the target values, then calculates the gradient of the output layer. This process is repeated recursively to compute the loss and gradients for each preceding layer in the network.}
        \label{fig:loss}
    \end{subfigure}
    
    \caption{The structure of the ANN.}
    \label{fig:3_ann}
\end{figure}

Activation functions \citep{ramachandran2017searching} introduce nonlinearity into the network, enabling it to learn complex patterns. Common activation functions include the Sigmoid \citep{kyurkchiev2015sigmoid}, ReLU \citep{chen2020dynamic}, and Tanh \citep{fan2002travelling}. For example, given an input $x$, a weight $\omega_1$, and bias $b$, the activation output is:
\begin{equation}
h = f(\omega_1 x + b),
\end{equation}
where $f$ is the activation function. A basic structure of ANN showed as \ref{fig:ANN}, and the activation function shown as fig. \ref{fig:activition}.

\subsubsection{Backpropagation and Optimization}

Neural networks are trained using the backpropagation algorithm \citep{hecht1992theory}, which adjusts weights by minimizing a loss function via gradient descent. A common loss function is the Mean Squared Error:
\begin{equation}
L = \frac{1}{N}\sum^N_{i=1} (y_i - \hat{y}_i)^2.
\end{equation}
The weights are updated according to:
\begin{equation}
x = x - \eta\frac{\partial L}{\partial x},
\end{equation}
where $\eta$ is the learning rate. Selecting an appropriate learning rate is crucial for efficient convergence \citep{he2019control}. Fig. \ref{fig:loss} shows how back propagation works.

\subsubsection{Convolutional Neural Networks (CNNs)}

\begin{figure}[h]
	\centering
	\includegraphics[width=0.7\textwidth]{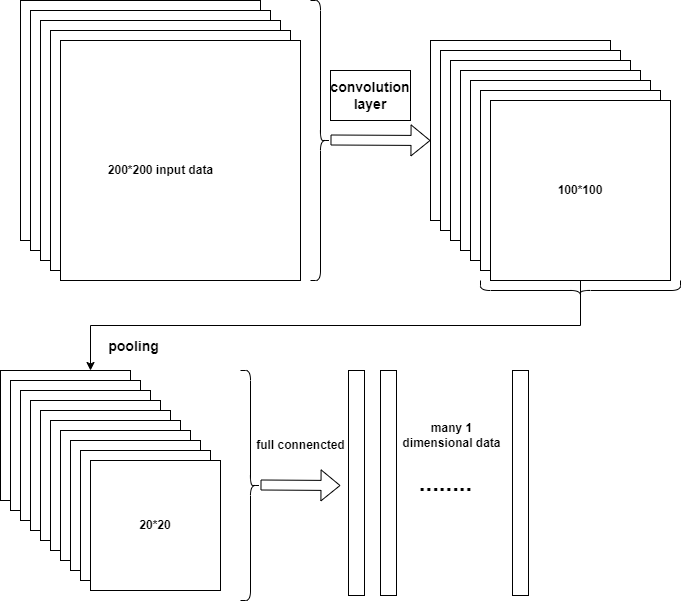}
	\caption{The basic architecture of CNN. Through convolution and pooling layers, the two-dimensional input data is progressively reduced in size using different operations. Finally, a fully-connected layer transforms the two-dimensional features into one-dimensional output.}
	\label{fig:CNN}
\end{figure}

In this work, we employ Convolutional Neural Networks (CNNs) \citep{yamashita2018convolutional}, which are particularly well-suited for image processing tasks due to their ability to exploit spatial hierarchies in data.  A basic structure of the CNN shown in Fig. \ref{fig:CNN}. 

CNNs typically consist of convolutional layers, pooling layers, and fully connected layers. The convolutional layer applies learnable filters (kernels) that slide over the input to extract local features \citep{cong2023review}. Pooling layers (e.g., max or average pooling) downsample the feature maps to reduce dimensionality while preserving key features \citep{shyam2021convolutional}. Finally, fully connected layers convert the processed feature maps into a flat vector for final prediction.

\subsubsection{Batch Size and Training}

We use large batch sizes in training process to facilitate computation of the output covariance matrix. Batch training allows for efficient parallel processing, especially on GPUs \citep{you2017large}. In each batch, a forward pass computes predictions, followed by a loss calculation and a backward pass to update weights. One training epoch completes after processing all batches.

\subsection{The Artificial Neural Networks in This Work}
\begin{figure*}
	\centering
	\includegraphics[width=1\textwidth]{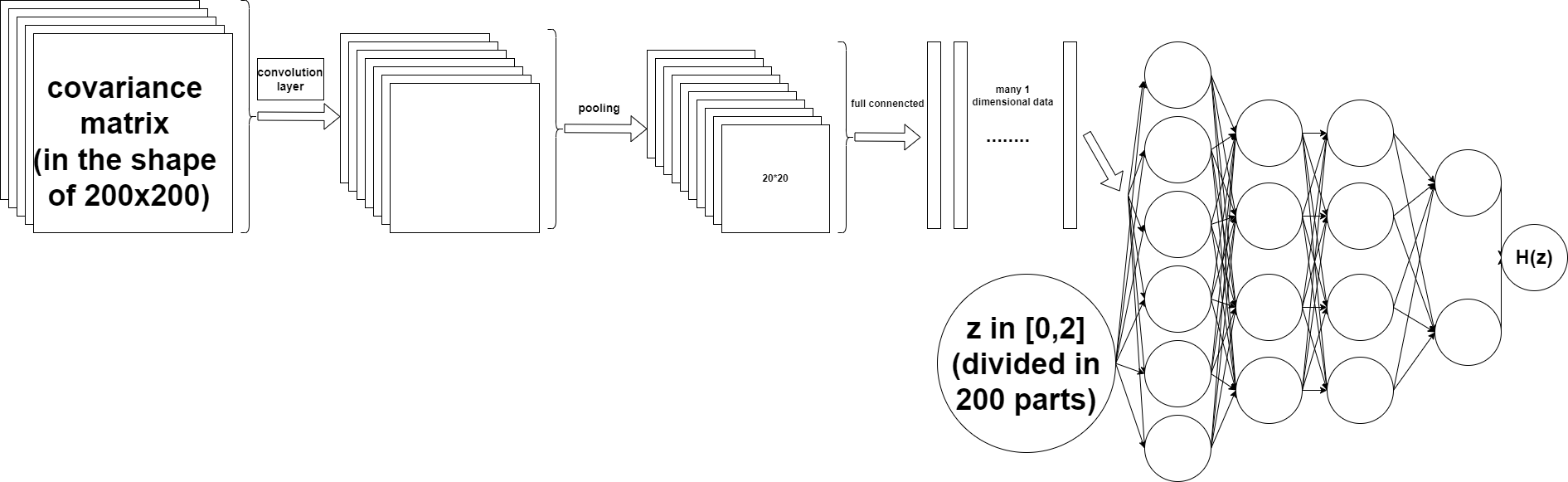}
	\caption{The network we used in our work. The network architecture integrates both convolutional and fully-connected components. Since the input covariance matrix is two-dimensional, we first process it through a CNN. The CNN part transforms this matrix into a one-dimensional feature vector, enabling concatenation with other one-dimensional input data for subsequent processing in the fully-connected network. In the right part, we built a normal full connected network. Besides the compressed covariance matrix, we also input the redshift in the range $[0,2]$ to the fully-connected network. Finally, the shape of the output is $(200,)$, representing the reconstructed $H(z)$ in the redshift range of $[0,2]$.}
	\label{fig:my}
\end{figure*}

In our work, we tried to reconstruct the $H(z)$ in the range of $z\in[0,2]$. As shown in the Fig. \ref{fig:my}, in the begining of our neural network, we built a CNN to process the two-dimensional covariance matrix. Then we added the z (redshift) to the network with the compressed covariance matrix to a full connected network. Flinally, the network outputs the reconstructed H(z) in redshift range $[0,2]$ as a 200-element array. In our work, to compute the covariance loss efficiently, we let the network directly output reconstructed data with a shape of $(1000, 200)$ (An example shown in Fig. \ref{fig:H_pred}). Here, the first dimension (1000) corresponds to 1000 distinct reconstruction of $H(z)$ in the redshift range of $z \in [0, 2]$ (note: this sample size can be increased as needed, e.g., to 2000 or much larger). Therefore the network would be simultaneous processing of 1000 covariance matrices, each with dimensions $(200, 200)$, resulting in a 3D tensor structure $(1000, 200, 200)$. Through convolutional layers, these are compressed to an output shape of $[1000, D_{\text{cov}}]$, where $D_{\text{cov}}$ denotes the reduced dimensionality of the compressed covariance representation. Consistent with this architecture, the redshift inputs are structured as a $(1000, 32)$ matrix, representing 1000 sets of redshift values paired with their corresponding $H(z)$ realizations.

In our neural network, the loss consists of two parts: the loss of the reconstruction and the loss of the covariance matrix. Firstly, we calculated the reconstruction loss between the output and the training data with MSE (mean squared error) loss:
\begin{equation}
Loss_{reconstruction}=\frac{1}{n}\sum^n_{i=1}(H(z)_i-\hat{H}(z)_i).
\end{equation}
Then we further calculated the covariance matrix of the output data, and thus computed the MSE loss between the training covariance matrix and the covariance matrix of the output data, in our work we define it as covariance loss:
\begin{equation}
Loss_{covariance}=\frac{1}{n}\sum^n_{i=1}(cov_i-\hat{cov}_i).
\end{equation}

\section{Data}
\label{sec:3}

\subsection{The OHD data We Used in This Work}

\begin{figure}[h]
	\centering
	\includegraphics[width=0.7\textwidth]{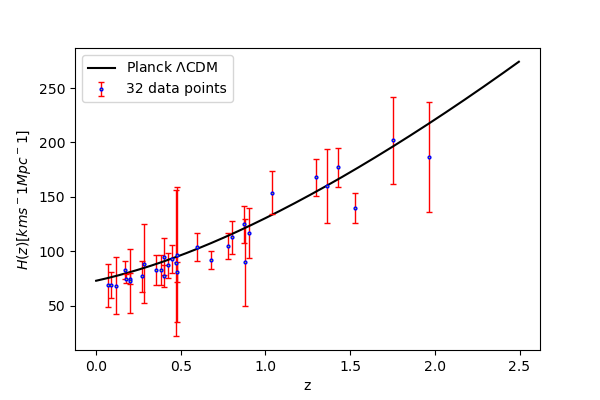}
	\caption{In this work, we employ 32 observational data points obtained through model-independent methods to avoid potential biases from theoretical assumptions. For comparison, we also include the $H(z)$ prediction from Planck Collaboration \citep{aghanim2018planck} with cosmological parameters $\Omega_\Lambda=0.686$, $\Omega_m=0.314$, and $H_0=67.4$ km s$^{-1}$ Mpc$^{-1}$. The Planck curve shows general agreement with the distribution of our data points.}
	\label{fig:32points}
\end{figure}

The real OHD is composed of $z_i,H(z_i)$ and $\sigma_i $, where $z_i$ is the redshift, and $H(z_i)$ is the corresponding Hubble parameter and $\sigma_i$ is the corresponding uncertainty. The 32 OHD data points we used in this work are evaluated with the cosmic chronometer method, which are given in \cite{Jimenez_2003}, \cite{PhysRevD.71.123001}, \cite{2009Cosmic}, \cite{2012New}, \cite{ratsimbazafy2017age}, \cite{cong2014four}, \cite{moresco20166}, \cite{jiao2023new} and \cite{10.1093/mnrasl/slv037}, and are shown in Fig. \ref{fig:32points}.

\subsection{The Mock OHD}

According to the flat $\Lambda\mathrm{CDM}$ model, the Hubble parameter is expressed by redshift $z$ with the simple formula:
\begin{equation}\label{eq11}
H(z) = H_{0}\sqrt{\Omega_{m}(1+z)^{3}+\Omega_{\Lambda}} ,
\end{equation}
where the $H_{0}$ is the Hubble constant. At the same time, the Hubble parameter can also be given by the non-flat $\Lambda\mathrm{CDM}$ model:
\begin{equation}\label{eq11_2}
H(z) = H_{0}\sqrt{\Omega_{m}(1+z)^{3}+\Omega_{\Lambda}+\Omega_{k}(1+z)^{2}} .
\end{equation}

\begin{figure}[h]
    \centering
    \begin{subfigure}{0.49\textwidth}
        \centering
        \includegraphics[width=\textwidth]{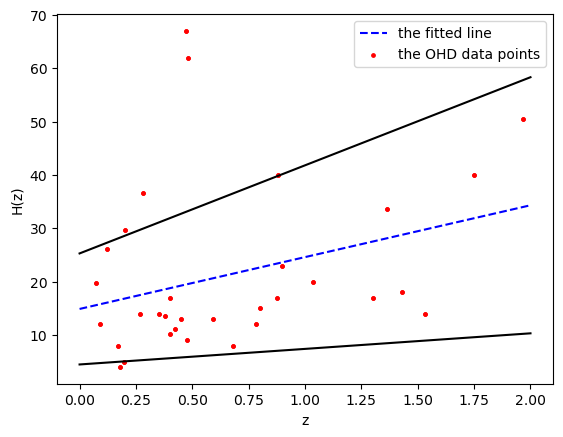} 
        \caption{The error model of the mock $H(z)$ \citep{ma2011power}. We sample the error ($\Delta H_i$) of the mock $H(z)$ in the region between two blue lines, the redshift range of $[0,2]$. The blue line means the fitted equation of the distribution of the error of the observational data. In this model, we gave up the two above points, for the reason that they are too away from the other data points.}
        \label{fig:3lines}
    \end{subfigure}
    \hfill 
    \begin{subfigure}{0.50\textwidth}
        \centering
        \includegraphics[width=\textwidth]{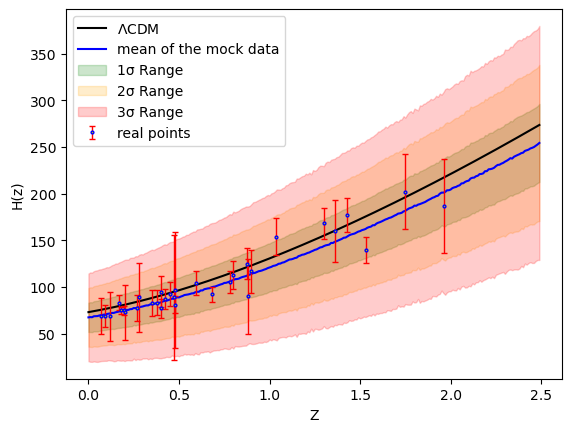}
        \caption{The $1 \sigma$, $2 \sigma$ and $3 \sigma$ of the mock $H(z)$. We had generated 10000 mock $H(z)$ points. Here we show the redshift in the range of [0,2.5], but the biggest $z$ of the OHD data points is just around 2, so in the work we just tried to reconstruct the $H(z)$ in the redshift range of $[0,2]$.}
        \label{fig:mockhz}
    \end{subfigure}
    
    \caption{The construction of the mock data.}
    \label{fig:training_set.}
\end{figure}

In our work, we generated the mock $H(z)$ with the flat $\Lambda\mathrm{CDM}$ model (Eq. \ref{eq11}), with the fiducial $H_0=67.4 km s^{-1}$Mpc$^{-1}$ and $\Omega_m = 0.314$ \citep{aghanim2018planck}. In the first place, we generate the error of the mock data. With the same error method used by \cite{ma2011power}, we assumed that the error of $H(z)$ increases linearly with the redshift. We firstly fitted $\sigma_{H(z)}$ with first degree polynomials and obtain $\sigma_0=9.72z+14.87$ (the blue dashed line). Here we assumed that $\sigma_0$ is the mean value of $\sigma_{H(z)}$ at a specific redshift. And then, two lines (the black solid lines) are selected symmetrically around the mean value line to ensure that most data points are in the area between them, and these two lines have the functions of $\sigma_\_=2.92z+4.46$ and $\sigma_+=16.52z+25.28$, which are shown in the Fig. \ref{fig:3lines}. Finally, the error $\sigma(z)$ was generated (or sampled) randomly according to the Gaussian distribution $\mathcal{N}(\sigma_0(z),\varepsilon(z))$, where $\varepsilon(z)=(\sigma_+ -\sigma_-)/4$ was set to ensure that $\sigma(z)$ falls in the area with a $95\%$ probability.

The fiducial values of the Hubble parameter $H_{\mathrm{fid}}(z)$ generated using Eq. \ref{eq11} are simulated randomly by adding $\Delta H$ subject to $\mathcal{N}(0,\varepsilon(z))$.  Thus, we can obtain the $H_{\mathrm{moc}}$ with the formula:
\begin{equation}\label{eq12}
H_{\mathrm{moc},i}=H_{\mathrm{fid}}(z_i)+\Delta H_i,
\end{equation}
where $z_i$ is in the redshift range $[0,2]$. We show the mock $H(z)$ in Fig. \ref{fig:mockhz}, we generated 10000 data points and plotted the $1 \sigma$, $2 \sigma$ and $3 \sigma$ of the distribution. In our work, the mock data serves two purposes. First, it is used to train the neural network. Second, we perform Bayesian inference with it, and the result will be a comparison to the result we got with the reconstruction data.

\subsection{The Training Covariance Matrix in This Work}

Firstly, we will briefly introduce the Gaussian process \citep{mackay1998introduction}, then we explain how we generate the covariance matrix with the Gaussian process regression \citep{williams1995gaussian}.

\subsubsection{The Gaussian Process Regression}

Gaussian Process Regression (GPR) is a non-parametric, probabilistic model that assumes the data is generated from a stochastic process. A Gaussian process (GP) is defined as a collection of random variables, any finite subset of which follows a multivariate normal distribution. In essence, GPR performs regression by modeling data as samples drawn from a GP, where the distribution over functions is specified by a mean function and a covariance (kernel) function.

A one-dimensional Gaussian distribution is characterized by its mean $\mu$ and variance $\sigma^2$:
\begin{equation}\label{eq1wei}
x \sim \mathcal{N}(\mu, \sigma^2),
\end{equation}
\begin{equation}
f(x) = \frac{1}{\sigma \sqrt{2\pi}} \exp\left(-\frac{(x - \mu)^2}{2\sigma^2}\right).
\end{equation}

For a multivariate Gaussian distribution, the distribution is defined by a mean vector and a covariance matrix $\Sigma$:
\begin{equation}
X \sim \mathcal{N}(\mu, \Sigma).
\end{equation}

A Gaussian process generalizes this concept to an infinite-dimensional case, defined over a continuous input space:
\begin{equation}
f(x) \sim \mathcal{GP}(\mu(x), k(x, x')),
\end{equation}
where $\mu(x)$ is the mean function, often assumed to be zero, and $k(x, x')$ is the covariance function (kernel) that encodes similarity between inputs.

In a regression setting, we model observations $y^{(i)}$ as:
\begin{equation}
y^{(i)} = f(x^{(i)}) + \epsilon^{(i)}, \quad \epsilon^{(i)} \sim \mathcal{N}(0, \sigma^2),
\end{equation}
where $f(x)$ is drawn from a GP.

Given training data $(X, Y)$ and test inputs $X\_$, the joint distribution of training and test outputs is:
\begin{equation}
\begin{bmatrix} f(X) \ f(X\_) \end{bmatrix} \sim \mathcal{N}\left(0, \begin{bmatrix} K(X,X) & K(X,X\_) \ K(X\_,X) & K(X\_,X\_) \end{bmatrix} \right).
\end{equation}

Applying Bayes theorem yields the posterior predictive distribution:
\begin{align}
\mu_* &= K_* (K + \sigma^2 I)^{-1} Y, \\
\Sigma_* &= K\_{} - K_* (K + \sigma^2 I)^{-1} K\_^\top,
\end{align}
where $K = K(X,X)$, $K\_ = K(X\_*,X)$, and $K\_{} = K(X\_,X\_)$.

The marginal likelihood of the observations is:
\begin{equation}
p(y|x) = \frac{1}{\sqrt{(2\pi)^n |K + \sigma^2 I|}} \exp\left( -\frac{1}{2} y^\top (K + \sigma^2 I)^{-1} y \right),
\end{equation}
and the corresponding log-marginal likelihood is:
\begin{equation}
\log p(y|x, \theta) = -\frac{1}{2} y^\top (K + \sigma^2 I)^{-1} y - \frac{1}{2} \log |K + \sigma^2 I| - \frac{n}{2} \log 2\pi,
\end{equation}
which is typically maximized with respect to the kernel hyperparameters $\theta$.

A commonly used kernel is the Radial Basis Function (RBF) kernel, also known as the Gaussian kernel \citep{buhmann2000radial}:
\begin{equation}
k(x_i, x_j) = \sigma_f^2 \exp\left(-\frac{|x_i - x_j|^2}{2l^2}\right),
\end{equation}
where $l$ is the length-scale and $\sigma_f^2$ is the signal variance. The choice of kernel is crucial, as it controls the properties of the GP and the function space being modeled \citep{schulz2018tutorial}.

\subsubsection{GPR with The Python Package Scikit-learn}\label{GPR}

In our work, we used the Python package scikit-learn \citep{pedregosa2011scikit} to perform Gaussian Process Regression. Scikit-learn is a widely-used open-source machine learning library built on top of NumPy \citep{oliphant2006guide}, SciPy \citep{virtanen2020scipy}, and matplotlib \citep{tosi2009matplotlib}.

We employed the RBF kernel \citep{buhmann2000radial}, which is the default and most commonly used kernel for GPR. It is well-suited for modeling smooth, continuous functions and provides a good balance between model flexibility and generalization performance.

\subsubsection{The Generation of The Covariance Matrix}

In Gaussian process regression, the covariance matrix encodes the correlations between both observed and reconstructed points. Even when the number of reconstruction points exceeds the number of observations, the predictive distribution is fully determined by the kernel-induced covariance structure, which governs both the mean prediction and the associated uncertainties. Therefore, we generated the covariance matrix with the GP reconstruction. There are some specifically key steps of the process of calculating the covariance matrix:

\textbf{(1)}. Sampling 1000 sets of $H(z)$ in every specific redshift from the Gaussian process reconstruction $\mathcal{N}(H(z),\sigma_{H(z)})$. In other word, firstly we will generate an array in shape $(200,1000)$ (Because we have 200 redshift in the range $z\in[0,2]$.).

\textbf{(2)}.  For two Hubble parameters at the redshift $z_1$ and $z_2$, the covariance between them can be calculated by comparing the 1000 $H(z_1)$ and 1000 $H(z_2)$ values using the equation:
\begin{equation}
Cov(H(z_i),H(z_j))
=\frac{1}{N}\sum^N_{k=1}[(H(z_i)_k-\Bar{H}(z_i))(H(z_j)_k-\Bar{H}(z_j))],
\end{equation}
where $N = 1000$, which means the number of data points at the $H(z_1)$ and $H(z_2)$, and $\Bar{H}$$(z)$ means the average value of the 1000 $H(z)$ data points.

\textbf{(3)}.  Using the method of \textbf{(2)}, we can calculate the covariance between any two Hubble parameters with different redshifts. Therefore we can calculate the whole covariance matrix.

\begin{figure}[h]
	\centering
	\includegraphics[width=0.5\textwidth]{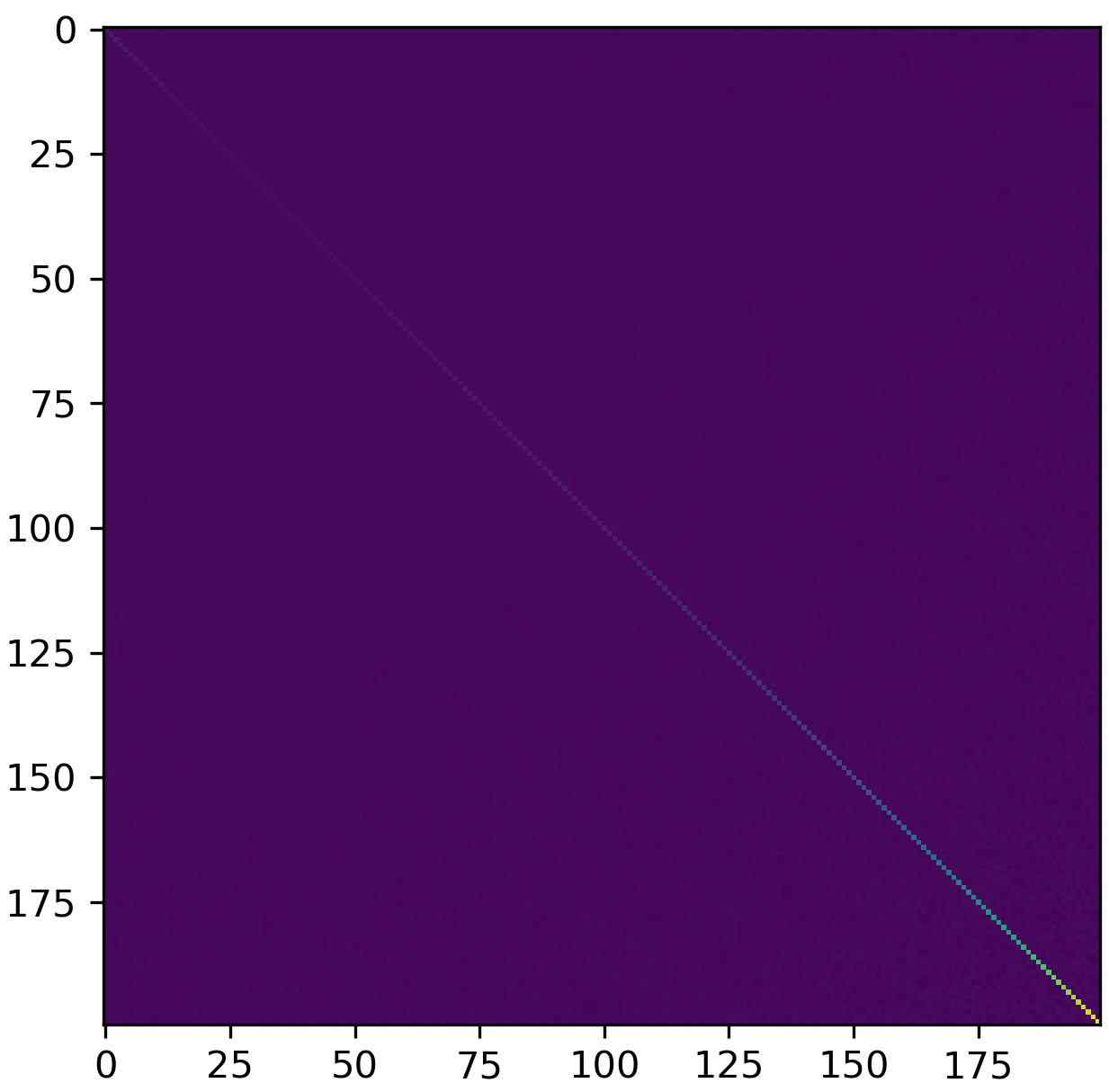}
	\caption{The full covariance matrix that generate with the $H(z)$ generated with the Gaussian process regression.}
	\label{fig:cov-quan}
\end{figure}

\begin{figure}[h]
    \centering
    \begin{subfigure}{0.5\textwidth}
        \centering
        \includegraphics[width=\textwidth]{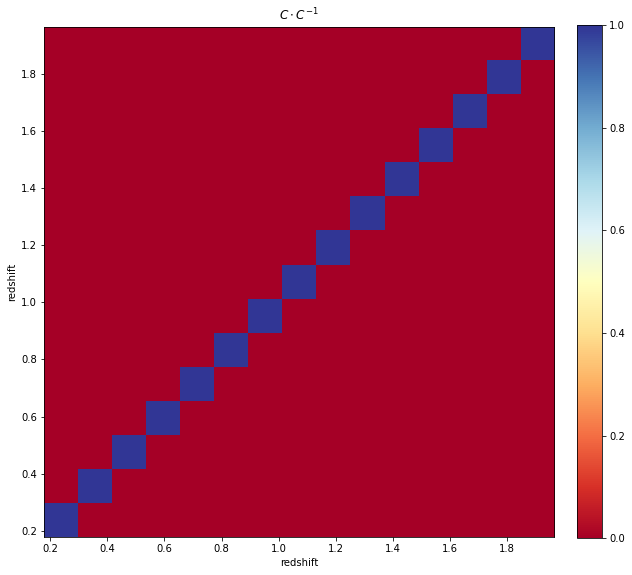} 
        \caption{The covariance matrix that generate by \cite{moresco2020setting}. There are only values on the diagonal line.}
        \label{fig:cov_jiaokang}
    \end{subfigure}
    \hfill 
    \begin{subfigure}{0.45\textwidth}
        \centering
        \includegraphics[width=\textwidth]{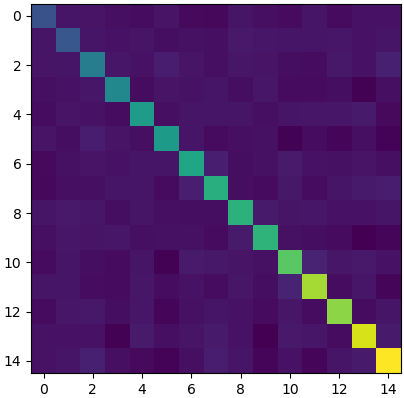}
        \caption{We randomly took 15 data points from the generated covariance matrix.}
        \label{fig:cov-15}
    \end{subfigure}
    
    \caption{The comparison of the covariance matrix.}
    \label{fig:2variances.}
\end{figure}

We draw the figure of the covariance matrix in Fig. \ref{fig:cov-quan} and randomly chose 15 data points to draw the heatmap in Fig. \ref{fig:cov-15}. Here we quoted the covariance matrix calculated by \cite{moresco2020setting} (we show it in the Fig. \ref{fig:cov_jiaokang}), which represents the covariance of the $H(z)$ in the redshift region of $[0,2]$. The covariance matrix calculated by \cite{moresco2020setting} and the covariance matrix calculated with the GPR share similar characteristic: almost only having values on the diagonal line. We can clearly see that the covariance matrix of the mock $H(z)$ is consistent with the result from \cite{moresco2020setting}, almost only having values on the diagonal line as well. In \cite{G_mez_Vargas1_2023}, the covariance matrix they generated also only has value on the diagonal line. The purpose of our work is to reconstruct OHD data in the redshift range of $[0,2]$ based on the real OHD data, and we hope the reconstructed data could be used as real data, so the reconstructed covariance matrices should be consistent with the real OHD data. It might be better to avoid adding too many additional components.

\section{Reconstruction and Parameters Estimation}
\label{sec:4}

\subsection{MCMC Method}

Firstly, in this subsection, we briefly introduce the MCMC method (Markov chain Monte Carlo, \citep{lewis2002cosmological,christensen2001bayesian}), which is the method we used to calculate the posterior distribution of the cosmological parameters. 

 MCMC method is widely used in various fields, including Bayesian inference \citep{yuan2015breaking}, physics, and machine learning \citep{andrieu2003introduction}. The process of MCMC involves simulating a Markov chain \citep{norris1998markov} in the parameter space, sampling randomly and then satisfying the samples to the target probability distribution function (PDF). It proceeds as follows: 

\textbf{(1)}. Initialization: Start from a random arbitrary initial state $x_0$.

\textbf{(2)}. Proposal: Propose a new state $x'$ based on a proposal distribution $Q(x'|x_t)$, which defines the probability of transitioning from state $x_t$ to $x'$.

\textbf{(3)}. Acceptance: Calculate the acceptance probability $(\alpha)$ to decide whether to accept or reject the proposed state.
\begin{equation}
\alpha (x_t,x')= min\left(1,\dfrac{P(x')*Q(x_t|x')}{P(x_t)*Q(x'|x_t)}\right)
\end{equation}

If $\alpha>1$ accept $x'$ as the next state.

If $\alpha<1$ accept $x'$  with probability $\alpha$ and stay at $x_t$ otherwise.
	
\textbf{(4)}. Repeat: Generate a series of states by iterating steps 2 and 3.

This process generates a sequence of states that eventually converge to samples from the target distribution $P(x)$. The samples can then be used to estimate expectations or calculate integrals over the target distribution.

MCMC method has many different kinds of algorithms, such as the Metropolis-Hastings \citep{chib1995understanding}, which provides a way to explore complex probability distributions, allowing researchers to draw samples from distributions that are otherwise challenging to sample directly.

The MCMC method we used in our work is Emcee \citep{foreman2013emcee}. Emcee is a Python package for performing Markov Chain Monte Carlo (MCMC) simulations, based on the affine-invariant ensemble sampler proposed by \citet{goodman2010ensemble}, rather than the traditional Metropolis-Hastings algorithm. It is particularly well-suited for Bayesian inference in high-dimensional and highly correlated parameter spaces. By using an ensemble of walkers that collectively explore the posterior distribution, emcee provides an efficient and user-friendly framework widely adopted in fields such as astrophysics, statistics, and machine learning \citep{box2011bayesian}. Like other MCMC methods, one advantage Emcee has is that it generates 10000 (or we can define the number by ourselves) data points with Markov chain in the end to represent the posterior distribution, at the same time it also provides the uncertainties.

\subsection{Reconstruction of $H(z)$}

\begin{figure}[h]
    \centering
    \begin{subfigure}{0.45\textwidth} 
        \centering
        \includegraphics[width=\textwidth]{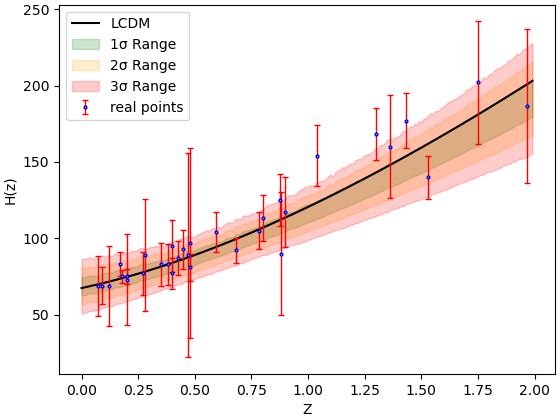} 
        \caption{The reconstructed $H(z)$, and the $1 \sigma$, $2 \sigma$ and $3 \sigma$. This result looks decent, and is quite similar to the Fig. \ref{fig:mockhz}. It basically matches the $\Lambda CDM$. The shape is smooth.}
        \label{fig:H_pred}
    \end{subfigure}
    
    \vspace{0.5cm} 
    \begin{subfigure}{0.45\textwidth}
        \centering
        \includegraphics[width=\textwidth]{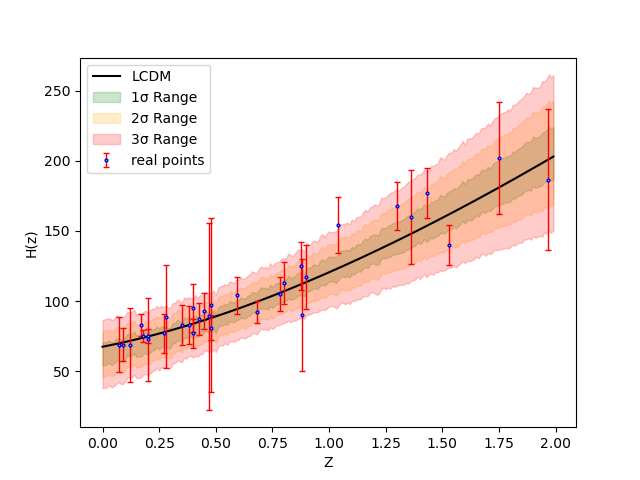}
        \caption{The reconstruction with loss dominates. The edge is not smooth.}
        \label{fig:badHrec2}
    \end{subfigure}
    \hfill
    \begin{subfigure}{0.45\textwidth}
        \centering
        \includegraphics[width=\textwidth]{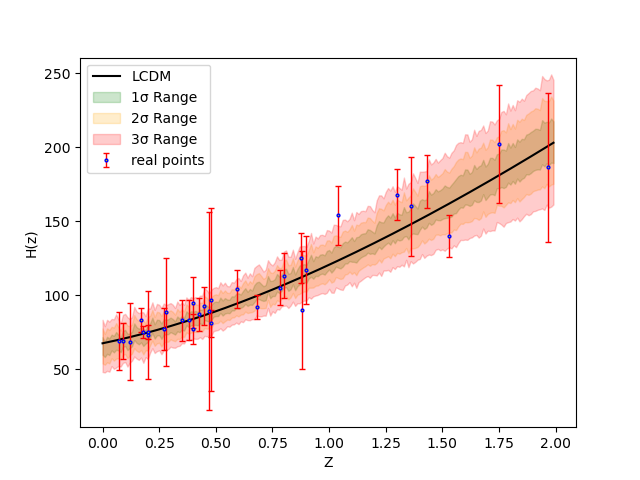}
        \caption{The reconstructed $H(z)$ reconstructed without covariance matrix. The shape of the $H(z)$ is mucher rougher.}
        \label{fig:Hrec_without_cov}
    \end{subfigure}
    
    \caption{Different kinds of reconstructions.}
    \label{fig:4_reconstruction.}
\end{figure}

In the Fig. \ref{fig:H_pred}, we show the reconstruction of $H(z)$ generated with our neural network. Besides, we show other 2 different reconstructed datasets. Fig. \ref{fig:badHrec2} was generated when the reconstructed dominates, Though it has a normal range, the edge is rough.

\begin{figure}[h]
    \centering
    \begin{subfigure}{0.48\textwidth} 
        \centering
        \includegraphics[width=\textwidth]{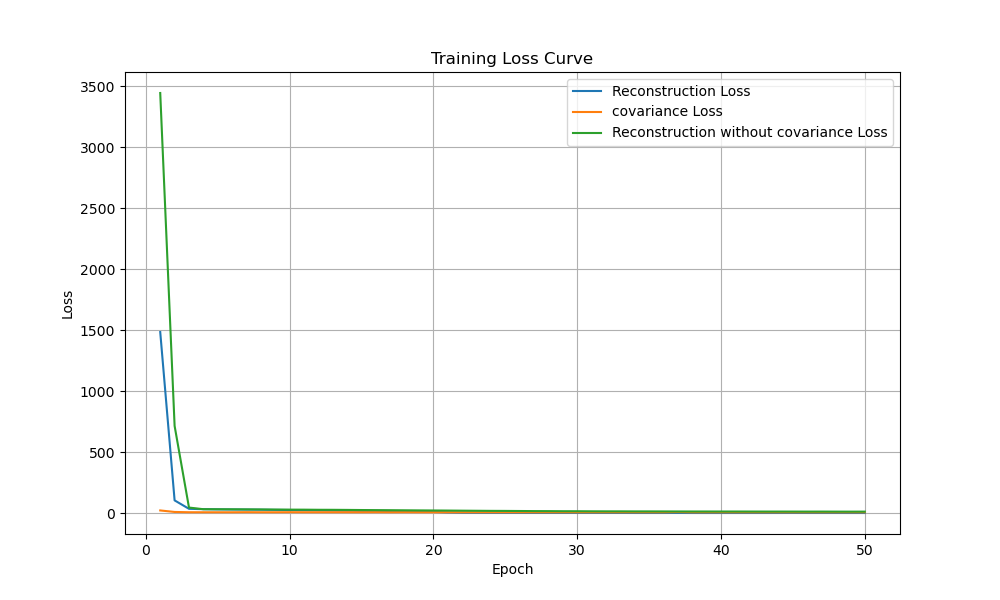} 
        \caption{The training loss curve of the reconstruction loss, covariance loss, and the reconstruction loss whitout covariance loss. They decrease rapidly and it is hard to see the difference, we show the detail in the following 3 figures with different methods.}
        \label{fig:3training_loss_curve}
    \end{subfigure}
    \hfill 
    \begin{subfigure}{0.48\textwidth}
        \centering
        \includegraphics[width=\textwidth]{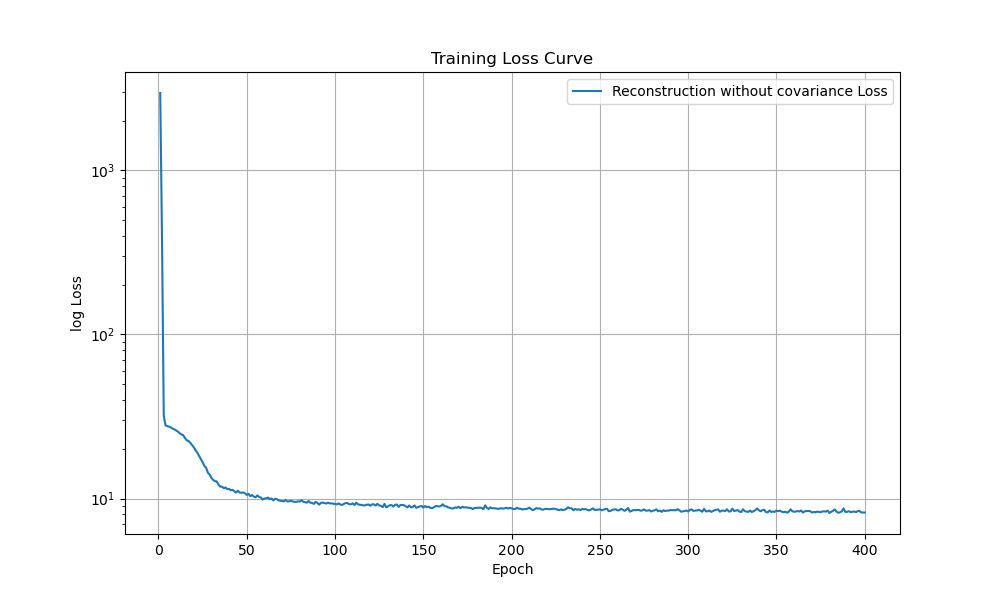}
        \caption{The reconstruction loss whitout covariance loss, we expand the x-axis up to 400. We can see that it took more epochs to obtain a minimum value.}
        \label{fig:training_loss_curve_only}
    \end{subfigure}
    
    \vspace{0.5cm} 
    \begin{subfigure}{0.48\textwidth}
        \centering
        \includegraphics[width=\textwidth]{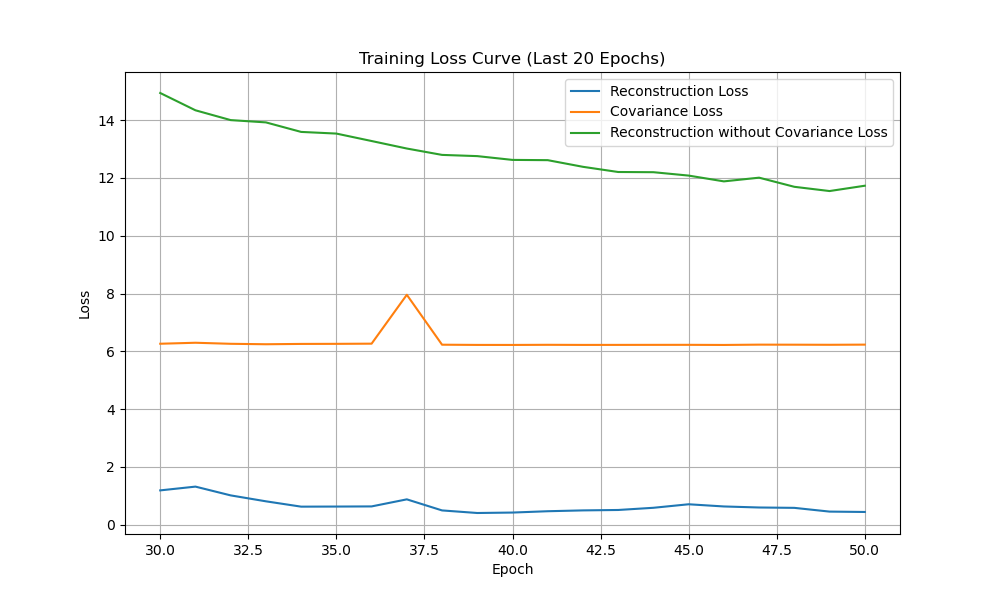}
        \caption{We show the last 20 epochs of the 3 loss, we can see that the the reconstruction loss whitout covariance loss still has a relatively large value.}
        \label{fig:training_last_20_epochs}
    \end{subfigure}
    \hfill
    \begin{subfigure}{0.48\textwidth}
        \centering
        \includegraphics[width=\textwidth]{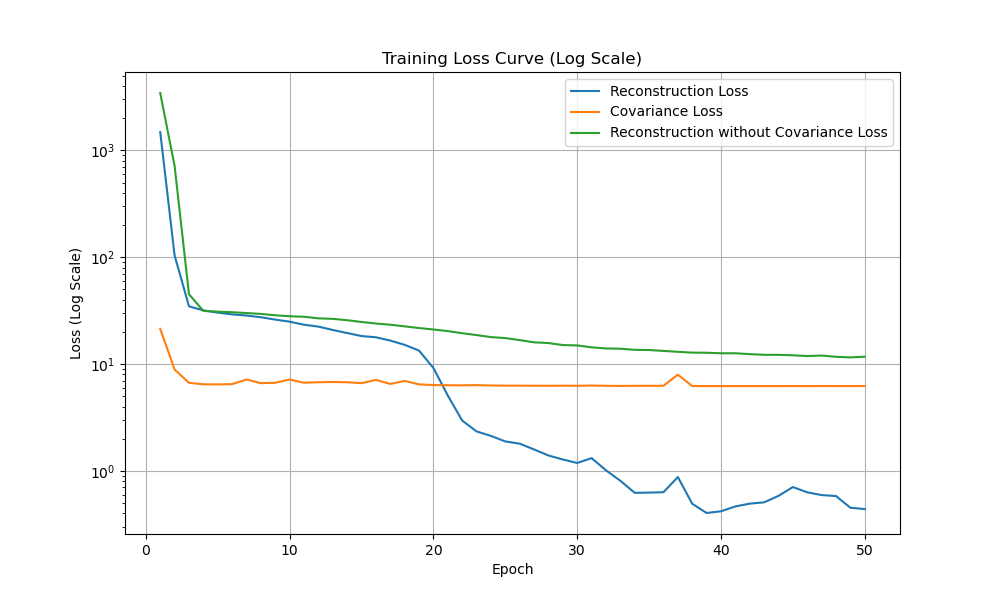}
        \caption{We use the Log Scale Curve. We can clearly see that the reconstruction loss whitout covariance loss  decreases slowly comparing to the other 2 losses.}
        \label{fig:Log_Scale_training_loss}
    \end{subfigure}
    
    \caption{The analysis of the training loss curve.}
    \label{fig:4_training_loss}
\end{figure}

We also test the situation that reconstructing the $H(z)$ without covariance loss, showing the result in the Fig. \ref{fig:Hrec_without_cov}. Without the covariance loss, the edge is much rougher, and it took a longer time to train the network. We drew the training loss curve in the Figs. \ref{fig:4_training_loss}. Fig. \ref{fig:3training_loss_curve} shows the reconstruction loss, the covaricance loss, and the rencostruction loss when we didn't use covariance, all of them decrease rapidly. However, if we expand the epoch and only see the reconstrcution loss (shown in Fig. \ref{fig:training_loss_curve_only}), we find that it takes more epochs to reach minimum value. Even in the Fig. \ref{fig:training_loss_curve_only}, the loss was still decreasing and didn't reach the minimum value. We use different methods to see the trend of the losses in Fig. \ref{fig:training_last_20_epochs} and Fig. \ref{fig:Log_Scale_training_loss}, both of them show that the reconstruction loss decreases slowly without covariance loss. Therefore, the covariance loss is necessary.

\subsection{Reconstruction of The Hubble Constant}

We also calculate the Hubble constant from the reconstructed $H(z)$. The Hubble constant is an important parameter in cosmology. According to the Friedman equations \citep{friedman1922krummung}:
\begin{equation}
\dot{a}^2-\frac{1}{3}(8\pi G\rho+\Lambda)a^2=-kc^2,
\end{equation}
and
\begin{equation}
\frac{\ddot{a}}{a}=-\frac{4}{3}\pi G(\rho+3\frac{p}{c^2})+\frac{1}{3}\Lambda.
\end{equation}
Here $a=a(t)$ is the scale factor of the Universe. At any given time, we can define a Hubble parameter:
\begin{equation}
H(t)=\frac{\dot{a}}{a}.
\end{equation}
It is obvious that the Hubble constant $H_0$ is the value of $H$ at the current time. In our reconstruction $H(z)$, we could obtain the $H_0$ at $z = 0$, which gave us $H_0 = 68.67\pm 5.903$ km s$^{-1}$ Mpc$^{-1}$. With CMB-based measurements \citep{blake2011wigglez, beutler20116df}, it gives values of $H_0 = 67\pm$3.2 km s$^{-1}$ Mpc$^{-1}$, to 4.8 per cent precision. The Hubble constant from our reconstructed $H_0$ is closed to the one from CMB-based calculation.

\subsection{Estimation of the Cosmological Parameters with Reconstructed $H(z)$}

We sampled 32 data points randomly from the reconstructed $H(z)$ and applied them to constrain the cosmological parameters with MCMC method. The posterior distribution estimated by MCMC, which is shown in the Fig. \ref{fig:poster_re}, gives the result of $H_0=68.03^{+3.51}_{-3.58}$ km s$^{-1}$ Mpc$^{-1}$, $\Omega_{m}=0.27^{+0.05}_{-0.05}$, $\Omega_{\Lambda}=0.73^{+0.05}_{-0.05}$. 

\subsection{Estimation of the Cosmological Parameters with Mock $H(z)$}

At the same time, in oder to do a better comparison, we also sampled 32 data points randomly from the mock $H(z)$ and applied to constrain the cosmological parameters with MCMC method. The posterior distribution estimated by MCMC, which is shown in the Fig. \ref{fig:training data poster}, gives the result of $H_0=62.04^{+15.98}_{-13.7}$ km s$^{-1}$ Mpc$^{-1}$, $\Omega_{m}=0.38^{+0.30}_{-0.17}$, $\Omega_{\Lambda}=0.62^{+0.17}_{-0.30}$.

\subsection{Comparision and Analysis of the Results}

\begin{figure}[h]
    \centering
    \begin{subfigure}{0.48\textwidth} 
        \centering
        \includegraphics[width=\textwidth]{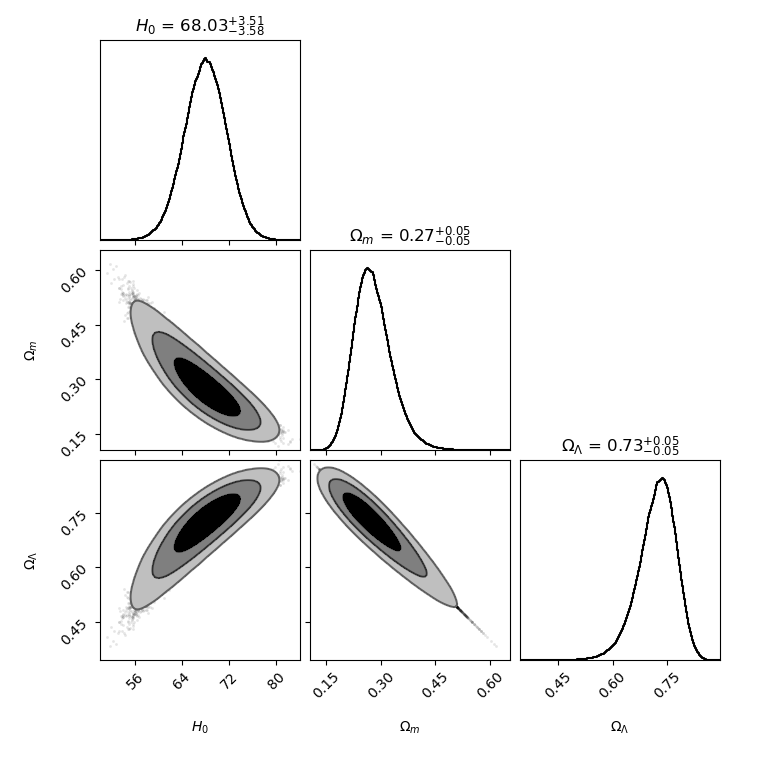} 
        \caption{Posterior distributions for 32 randomly sampled points from the reconstructed $H(z)$, derived with MCMC.}
        \label{fig:poster_re}
    \end{subfigure}
    \hfill 
    \begin{subfigure}{0.48\textwidth}
        \centering
        \includegraphics[width=\textwidth]{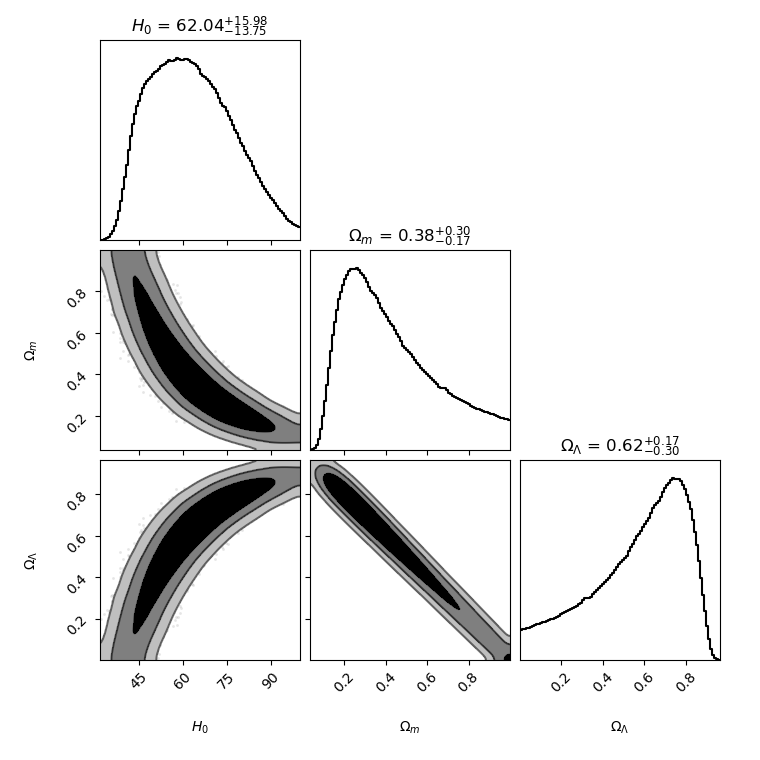}
        \caption{MCMC-computed posterior distributions of 32 randomly selected data points from the mock $H(z)$.}
        \label{fig:training data poster}
    \end{subfigure}
    
    \vspace{0.5cm} 
    \begin{subfigure}{0.48\textwidth}
        \centering
        \includegraphics[width=\textwidth]{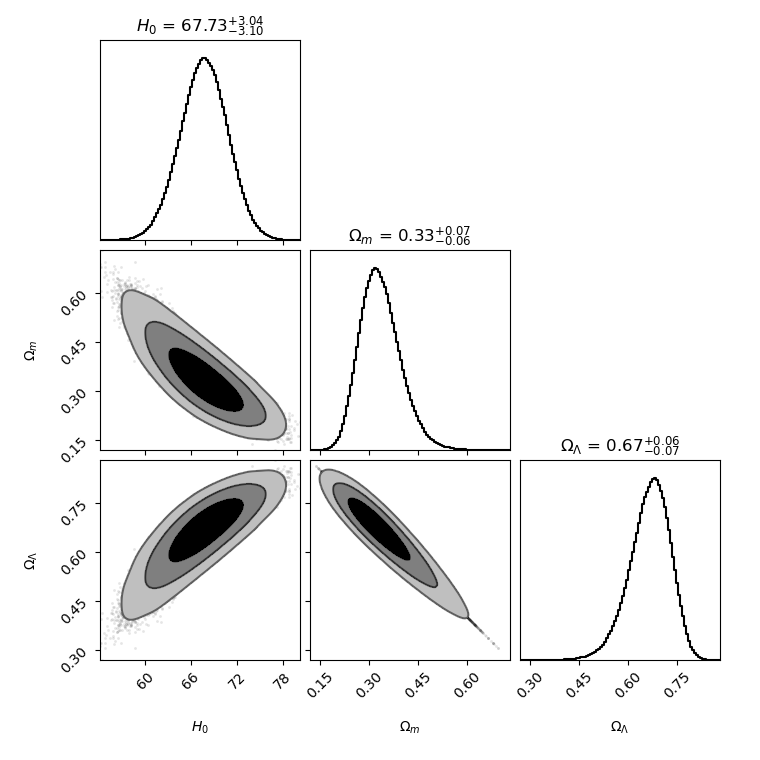}
        \caption{Posterior distributions for 32 OHD data points obtained via MCMC.}
        \label{fig:poster_rea}
    \end{subfigure}
    
    \caption{MCMC-derived posterior distributions for: (1) the reconstructed data (Fig. \ref{fig:poster_re}), (2) mock data (Fig. \ref{fig:training data poster}), and (3) real observational data (Fig. \ref{fig:poster_rea}).}
    \label{fig:3_distribution}
\end{figure}

Besides, we also calculated the posterior distribution with the real 32 OHD points, we show the result in the Fig. \ref{fig:poster_rea}. The real 32 OHD points gives the result of $H_0=67.73^{+3.04}_{-3.10}$ km s$^{-1}$ Mpc$^{-1}$, $\Omega_{m}=0.33^{+0.07}_{-0.06}$, $\Omega_{\Lambda}=0.73^{+0.06}_{-0.07}$.

It is obvious that the posterior distribution estimated with the data points from the reconstructed $H(z)$ is more closer to the roughly estimated values ($H_0\approx67 km s^{-1} Mpc^{-1}, \Omega_{\Lambda}\approx 0.7, \Omega_m\approx 0.3$), or more precisely, the $\Lambda$CDM model from Planck Collaboration \citep{aghanim2018planck}, which gives the estimation of $H_0\approx67.4 km s^{-1} Mpc^{-1}, \Omega_{\Lambda}\approx 0.686, \Omega_m\approx 0.314$. At the same time, the confidence interval of the posterior distribution estimated with the data points from the reconstructed $H(z)$ is tighter. 

Furthermore, in order to test the quality of the reconstructed $H(z)$, we apply two criteria, KL divergence and figure of merit(FoM).

\subsubsection{Comparison using KL divergence}

\emph{Kullback-Leibler divergence (KL divergence)}. Kullback–Leibler divergence is a statistical feature which can measure how one probability distribution is different from another one. With this characteristic, Kullback–Leibler divergence is widely used to calculate how much information is lost when we try to approximate one distribution with the second one.

While processing probability and statistics, we generally can simulate the observed data or complex distribution with a simpler approximate distribution, which benefits our following experiment. Suppose that we have two probability density distributions $p_1(\boldsymbol{\theta})$ and $p_2(\boldsymbol{\theta})$, where  $p_2(\boldsymbol{\theta})$ is the simulation of the $p_1(\boldsymbol{\theta})$. In this case, The KL divergence from $p_1(\boldsymbol{\theta})$ to $p_2(\boldsymbol{\theta})$ is defined as
\begin{equation}
D_{\mathrm{KL}}=(p_1(\boldsymbol{\theta})||p_2(\boldsymbol{\theta}))=\mathbb{E}_{p_1(\boldsymbol{\theta})}(\log\,p_1(\boldsymbol{\theta})-\log\,p_2(\boldsymbol{\theta})).
\end{equation}
In our work, we sample $M$ samples $\{\boldsymbol{\theta}_i \}$ from the posterior, so the KL divergence is estimated with:
\begin{equation}\label{eqkl}
D_{\mathrm{KL}}(p_1||p_2)=\frac{1}{M}\sum_{i=1}^{M}(\mathrm{ln}\,P_1(\boldsymbol{\theta}_i|\textbf{\emph{H}}_{\mathrm{obs}})-\mathrm{ln}\,P_2(\boldsymbol{\theta}_i|\textbf{\emph{H}}_{\mathrm{obs}})),
\end{equation} 
where $P_2(\boldsymbol{\theta}|\textbf{\emph{H}}_{\mathrm{obs}})$ is the posterior calculated with reconstructed $H(z)$ and $P_1(\boldsymbol{\theta}|\textbf{\emph{H}}_{\mathrm{obs}})$ is the posterior calculated with real 32 OHD data points. From Eq. (\ref{eqkl}), it is obvious that the smaller the KL divergence is, the closer the $P_1(\boldsymbol{\boldsymbol{\theta}}|\textbf{\emph{H}}_{\mathrm{moc}})$ and $P_2(\boldsymbol{\theta}|\textbf{\emph{H}}_{\mathrm{moc}})$ will be. When $D_{\mathrm{KL}}(p_1||p_2)=0$, it means that the two posterior are almost identical.

In our work, we employed KL divergence to compare the posterior distributions of the cosmological parameters  cosmological parameters $\Omega_{m}, \Omega_{\Lambda}$ and $H_{0}$, which are inferred from samples based on the reconstructed $H(z)$ and the observed Hubble data (OHD). We sampled 10, 20, 32, 64, 128, data points respectively from the reconstructed $H(z)$ and calculated their posterior distributions. Then we calculated the KL divergence comparing to the posterior distribution calculated by the real 32 data points. As the number of sampled data points increases, KL divergence decreases, which is reasonable because more data points generally means more accurate measurements. In Fig.\ref{fig:KLD}, we show the KL divergence measurement.

\subsubsection{Comparison using FoM}

\emph{Figure of merit}(FoM). The FoM is a statistical feature which can measure the range of the parameters and how tight the constraints are. The FoM used in this work is similar to the one adopted by \cite{ma2011power} and \cite{2011Constraints} in their work, which defined as:
\begin{equation}
P(\boldsymbol{\theta}|\textbf{\emph{H}}_{\mathrm{obs}})=\mathrm{const.}=\exp(-\Delta \mathcal{X}^2/2)P_{\mathrm{max}},
\end{equation}
where $P_{\mathrm{max}}$ is the maximum possibility density of the posterior, and $\exp(-\Delta \mathcal{X}^2/2)$ is a constant which ensures that $\exp(-\Delta \mathcal{X}^2/2)P_{\mathrm{max}}$ is equal to the probability density at the boundary of the 95.44\% confidence region of the Gaussian distribution. According to \cite{2021Likelihood}, we use the same $\exp(-\Delta \mathcal{X}^2/2)$ here, which is 8.02. The FoM means the reciprocal volume of the confidence region of the posterior distribution, so the larger the FoM is, the tighter the constraint of the parameters is.

\begin{figure}[h]
    \centering
    \begin{subfigure}{0.48\textwidth}
        \centering
        \includegraphics[width=\textwidth]{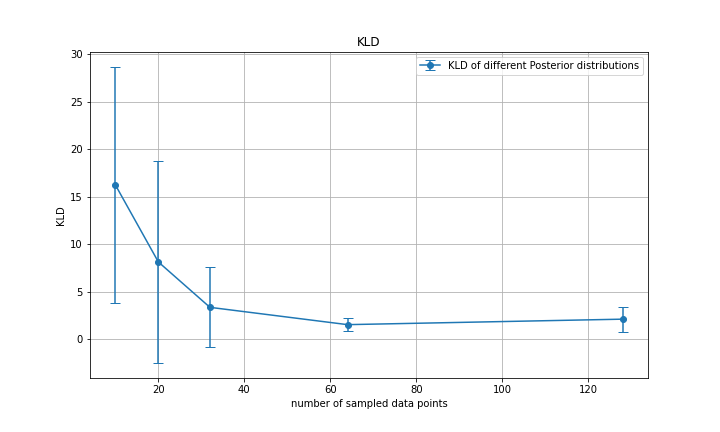} 
        \caption{KL divergence measurement. There is a obvious trend that when the number of the sampled data points increases, the KL divergence decreases, which is reasonable. In our work, we calculated the KL divergence based on the joint posterior distribution of the cosmological parameters $\Omega_{m}, \Omega_{\Lambda}$ and $H_{0}$.}
        \label{fig:KLD}
    \end{subfigure}
    \hfill 
    \begin{subfigure}{0.48\textwidth}
        \centering
        \includegraphics[width=\textwidth]{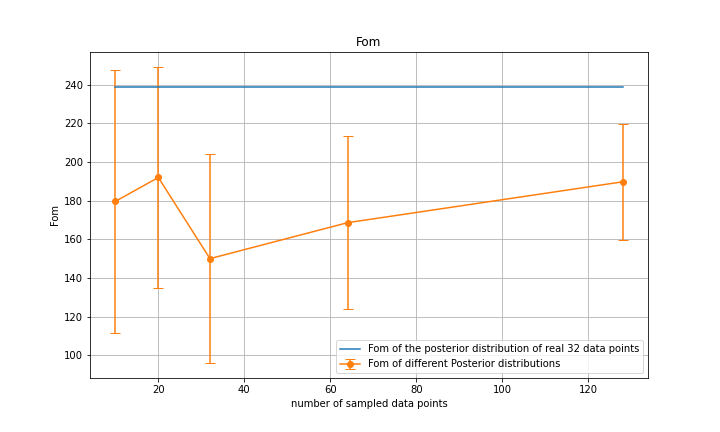}
        \caption{FoM measurement doesn't have obvious trend of change in different sampled data points, but the error decreases when the sample data points increases. In our work, we calculated the FoM based on the joint posterior distribution of the cosmological parameters $\Omega_{m}, \Omega_{\Lambda}$ and $H_{0}$.}
        \label{fig:FOM}
    \end{subfigure}
    
    \caption{The measurement of the KL divergence and FOM.}
    \label{fig:4_reconstruction.}
\end{figure}

In Fig.\ref{fig:FOM}, we show the FoM measurement. Unlike KLD measurement in Fig.\ref{fig:KLD}, there is no obvious trend of change in different sampled data points. But we can see that as the number of sampled data points increases, the error decreases, which is also reasonable because more data points will bring us sRle range of the posterior distribution. 

\section{Conclusions and Discussions}
\label{sec:5}

\subsection{Conclusions}

In our work, we had put forward a new method that can be used to reconstruct the $H(z)$. The effect of this method is obvious. One of the main purposes of our work is to estimate the cosmological parameters, so we sampled some data points from the reconstructed $H(z)$ and constrainded in the $\Lambda CDM$ model. We found that the posterior distribution calculated by the sampled reconstructed data is consistent with the $\Lambda CDM$ model with Planck Collaboration \citep{aghanim2018planck}. The posterior distribution obtained from the reconstructed H(z) shows confidence intervals comparable in size to those derived from the real 32 OHD data points. Notably, both distributions yield consistent parameter ranges for the three cosmological parameters, and the shapes (Fig. \ref{fig:poster_re} and Fig. \ref{fig:poster_rea}) are similar. We further tested the posteriors by using KL divergence and FoM, and the reconstructed data behaved well in the test. On the contrary, the shape and the range of the posterior distribution calculated by the mock $H(z)$ didn't behave well. At the same time, with the reconstructed $H(z)$, we can easily get the Hubble constant $H_0$, and the result is consistent with the Hubble constant calculated by the CMB-based measurements.

We combined the real observational data and covariance matrix to generate reconstructed $H(z)$ in a specific redshift range in $z\in[0,2]$. More precisely, we hope the reconstructed $H(z)$ can learn the characteristics of both Fig. \ref{fig:mockhz} and GP. To some extent, it is still reasonable because when there has more data points, it is easlier to get a precise estimation. On the contrary, Fig. \ref{fig:mockhz} lose the precision in order to construct the balanced shape of the distribution. For this reason, in our work we decided to combine the 2 distributions in Fig. \ref{fig:mockhz} and GP. 

In summary, our analysis demonstrates that the reconstructed $H(z)$ reliably captures the underlying distribution of observational data. As an accurate reconstruction of $H(z)$, it enables the exploration of important cosmological implications, including tighter constraints on the expansion history, the nature of dark energy, and potential deviations from the standard $\Lambda\mathrm{CDM}$ paradigm.
\subsection{Discussions}

We can see that the Bayesian inference results from the reconstruction and the real data are quite 
similar, but it doesn't mean that there is no added value in the proposed method. The purpose of our work is to reconstruct OHD data in the redshift range of [0,2] to compensate for the sparse data available in the redshift range, so we hope the Bayesian inference results from the reconstruction and the real data are consistent. Meanwhile, our reconstruction essentially is an extension based on the existing data. In principle, the Bayesian results obtained from the reconstructed data should not outperform those derived from the real data, as that would imply that additional information has been introduced.

Admittedly, we should pay attention to the construction of the covariance matrix we used as training data in our network, and see if we could find a better method to generate the covariance matrix. In the bigining, we tried to reconstruct the $H(z)$ in the redshitf range of $[0,2.5]$. However, when we uesd the GPR, the data diverged in the redshift range of $[2,2.5]$ because of having no data points in that redshift range. Consequently, it is hard for us to calculate the covariance matrix and the posterior distribution. For this reason, we had to limite the redshift in the range of $[0,2]$. In \cite{G_mez_Vargas1_2023}, they did a similar work, using VAE to generate the covariance matrix. Unlike their work, we directly input the covariance matrix to the neural network to help to generate reconstruction data. In \cite{moresco2024measuringexpansionhistoryuniverse, moresco2018setting, moresco2020setting, moresco2012improved, moresco20166}, they rigorously computed the covariance matrix through a systematic approach that accounts for key astrophysical parameters, including metallicity, contamination, and Stellar Population Synthesis (SPS) model. While covariance matrix is reliable, it remains limited to a 15×15 dimensionality (shown in Fig. \ref{fig:cov_jiaokang}), potentially restricting its applicability. However, this method is still significant because it provides reliable result. In the future, the method could be combined with additional OHD data points from upcoming surveys, such as CSST and Euclid, to generate a precise and comprehensive covariance matrix, enabling us to generate more reliabe reconstruction.

Besides, $\Lambda CDM$ is relatively easy to constrain. In the current study, our goal is to demonstrate and validate the proposed methodology within the well-established $\Lambda CDM$ framework, which serves as a solid baseline. We consider the extension to other models such as CPL , to be a valuable direction for future work, and we plan to explore these in the following studies.

Meanwhile, ANN models usually need careful hyperparameter fine-tuning before the best performance can be achieved.  Therefore, future work should be focused on finding some other better models, such as VEA-GAN \citep{cemgil2020autoencoding,mukesh2022variational} and AVAE \citep{plumerault2021avae}. Exploring efficient hyperparameter tuning strategies, as systematically compared in \cite{G_mez_Vargas_2023}, may further enhance performance. We hope we can find more useful models in the future work.

\begin{acknowledgements}

We sincerely thank the anonymous reviewers for their thorough review and valuable feedback, which helped us clarify key points and improve the overall presentation of the paper. We thank Shiyu Li, Yunlong Li, Yu Hu, Changzhi Lu, Yuchen Wang, Shulei Cao, Jin Qin, Zhenzhao Tao, Zijian Wang, Xiaohang luan, Jing Niu, guangzai ye, zijian si and Kang Jiao for useful discussions and their kind help. This work is supported by National Key R\&D Program of China (2023YFB4503305), National SKA Program of China (2022SKA0110202), the China Manned Space Program with grant No. CMS-CSST-2025-A01, the National Natural Science Foundation of China (Grants No.12373109, 61802428) and China Scholarship Council (File No.2306040042).
\end{acknowledgements}











\bibliographystyle{apsrmp4-2}
\bibliography{apssamp}

\begin{thebibliography}{74}%
\makeatletter
\providecommand \@ifxundefined [1]{%
 \@ifx{#1\undefined}
}%
\providecommand \@ifnum [1]{%
 \ifnum #1\expandafter \@firstoftwo
 \else \expandafter \@secondoftwo
 \fi
}%
\providecommand \@ifx [1]{%
 \ifx #1\expandafter \@firstoftwo
 \else \expandafter \@secondoftwo
 \fi
}%
\providecommand \natexlab [1]{#1}%
\providecommand \enquote  [1]{``#1''}%
\providecommand \bibnamefont  [1]{#1}%
\providecommand \bibfnamefont [1]{#1}%
\providecommand \citenamefont [1]{#1}%
\providecommand \href@noop [0]{\@secondoftwo}%
\providecommand \href [0]{\begingroup \@sanitize@url \@href}%
\providecommand \@href[1]{\@@startlink{#1}\@@href}%
\providecommand \@@href[1]{\endgroup#1\@@endlink}%
\providecommand \@sanitize@url [0]{\catcode `\\12\catcode `\$12\catcode
  `\&12\catcode `\#12\catcode `\^12\catcode `\_12\catcode `\%12\relax}%
\providecommand \@@startlink[1]{}%
\providecommand \@@endlink[0]{}%
\providecommand \url  [0]{\begingroup\@sanitize@url \@url }%
\providecommand \@url [1]{\endgroup\@href {#1}{\urlprefix }}%
\providecommand \urlprefix  [0]{URL }%
\providecommand \Eprint [0]{\href }%
\providecommand \doibase [0]{https://doi.org/}%
\providecommand \selectlanguage [0]{\@gobble}%
\providecommand \bibinfo  [0]{\@secondoftwo}%
\providecommand \bibfield  [0]{\@secondoftwo}%
\providecommand \translation [1]{[#1]}%
\providecommand \BibitemOpen [0]{}%
\providecommand \bibitemStop [0]{}%
\providecommand \bibitemNoStop [0]{.\EOS\space}%
\providecommand \EOS [0]{\spacefactor3000\relax}%
\providecommand \BibitemShut  [1]{\csname bibitem#1\endcsname}%
\let\auto@bib@innerbib\@empty
\bibitem [{\citenamefont {Aghanim}\ \emph {et~al.}(2018)\citenamefont
  {Aghanim}, \citenamefont {Akrami}, \citenamefont {Ashdown}, \citenamefont
  {Aumont}, \citenamefont {Baccigalupi}, \citenamefont {Ballardini},
  \citenamefont {Banday}, \citenamefont {Barreiro}, \citenamefont {Bartolo},
  \citenamefont {Basak} \emph {et~al.}}]{aghanim2018planck}%
  \BibitemOpen
  \bibfield  {author} {\bibinfo {author} {\bibnamefont {Aghanim}, \bibfnamefont
  {N.}}, \bibinfo {author} {\bibfnamefont {Y.}~\bibnamefont {Akrami}}, \bibinfo
  {author} {\bibfnamefont {M.}~\bibnamefont {Ashdown}}, \bibinfo {author}
  {\bibfnamefont {J.}~\bibnamefont {Aumont}}, \bibinfo {author} {\bibfnamefont
  {C.}~\bibnamefont {Baccigalupi}}, \bibinfo {author} {\bibfnamefont
  {M.}~\bibnamefont {Ballardini}}, \bibinfo {author} {\bibfnamefont
  {A.}~\bibnamefont {Banday}}, \bibinfo {author} {\bibfnamefont
  {R.}~\bibnamefont {Barreiro}}, \bibinfo {author} {\bibfnamefont
  {N.}~\bibnamefont {Bartolo}}, \bibinfo {author} {\bibfnamefont
  {S.}~\bibnamefont {Basak}},  \emph {et~al.}} (\bibinfo {year} {2018}),\
  \href@noop {} {\bibinfo  {journal} {arXiv preprint arXiv:1807.06209}\
  }\BibitemShut {NoStop}%
\bibitem [{\citenamefont {Andrieu}\ \emph {et~al.}(2003)\citenamefont
  {Andrieu}, \citenamefont {De~Freitas}, \citenamefont {Doucet},\ and\
  \citenamefont {Jordan}}]{andrieu2003introduction}%
  \BibitemOpen
\bibfield  {journal} {  }\bibfield  {author} {\bibinfo {author} {\bibnamefont
  {Andrieu}, \bibfnamefont {C.}}, \bibinfo {author} {\bibfnamefont
  {N.}~\bibnamefont {De~Freitas}}, \bibinfo {author} {\bibfnamefont
  {A.}~\bibnamefont {Doucet}}, and\ \bibinfo {author} {\bibfnamefont {M.~I.}\
  \bibnamefont {Jordan}}} (\bibinfo {year} {2003}),\ \href@noop {} {\bibfield
  {journal} {\bibinfo  {journal} {Machine learning}\ }\textbf {\bibinfo
  {volume} {50}},\ \bibinfo {pages} {5}}\BibitemShut {NoStop}%
\bibitem [{\citenamefont {Beutler}\ \emph {et~al.}(2011)\citenamefont
  {Beutler}, \citenamefont {Blake}, \citenamefont {Colless}, \citenamefont
  {Jones}, \citenamefont {Staveley~Smith}, \citenamefont {Campbell},
  \citenamefont {Parker}, \citenamefont {Saunders},\ and\ \citenamefont
  {Watson}}]{beutler20116df}%
  \BibitemOpen
  \bibfield  {author} {\bibinfo {author} {\bibnamefont {Beutler}, \bibfnamefont
  {F.}}, \bibinfo {author} {\bibfnamefont {C.}~\bibnamefont {Blake}}, \bibinfo
  {author} {\bibfnamefont {M.}~\bibnamefont {Colless}}, \bibinfo {author}
  {\bibfnamefont {D.~H.}\ \bibnamefont {Jones}}, \bibinfo {author}
  {\bibfnamefont {L.}~\bibnamefont {Staveley~Smith}}, \bibinfo {author}
  {\bibfnamefont {L.}~\bibnamefont {Campbell}}, \bibinfo {author}
  {\bibfnamefont {Q.}~\bibnamefont {Parker}}, \bibinfo {author} {\bibfnamefont
  {W.}~\bibnamefont {Saunders}}, and\ \bibinfo {author} {\bibfnamefont
  {F.}~\bibnamefont {Watson}}} (\bibinfo {year} {2011}),\ \href@noop {}
  {\bibfield  {journal} {\bibinfo  {journal} {Monthly Notices of the Royal
  Astronomical Society}\ }\textbf {\bibinfo {volume} {416}}~(\bibinfo {number}
  {4}),\ \bibinfo {pages} {3017}}\BibitemShut {NoStop}%
\bibitem [{\citenamefont {Blake}\ \emph {et~al.}(2011)\citenamefont {Blake},
  \citenamefont {Kazin}, \citenamefont {Beutler}, \citenamefont {Davis},
  \citenamefont {Parkinson}, \citenamefont {Brough}, \citenamefont {Colless},
  \citenamefont {Contreras}, \citenamefont {Couch}, \citenamefont {Croom} \emph
  {et~al.}}]{blake2011wigglez}%
  \BibitemOpen
  \bibfield  {author} {\bibinfo {author} {\bibnamefont {Blake}, \bibfnamefont
  {C.}}, \bibinfo {author} {\bibfnamefont {E.~A.}\ \bibnamefont {Kazin}},
  \bibinfo {author} {\bibfnamefont {F.}~\bibnamefont {Beutler}}, \bibinfo
  {author} {\bibfnamefont {T.~M.}\ \bibnamefont {Davis}}, \bibinfo {author}
  {\bibfnamefont {D.}~\bibnamefont {Parkinson}}, \bibinfo {author}
  {\bibfnamefont {S.}~\bibnamefont {Brough}}, \bibinfo {author} {\bibfnamefont
  {M.}~\bibnamefont {Colless}}, \bibinfo {author} {\bibfnamefont
  {C.}~\bibnamefont {Contreras}}, \bibinfo {author} {\bibfnamefont
  {W.}~\bibnamefont {Couch}}, \bibinfo {author} {\bibfnamefont
  {S.}~\bibnamefont {Croom}},  \emph {et~al.}} (\bibinfo {year} {2011}),\
  \href@noop {} {\bibfield  {journal} {\bibinfo  {journal} {Monthly Notices of
  the Royal Astronomical Society}\ }\textbf {\bibinfo {volume} {418}}~(\bibinfo
  {number} {3}),\ \bibinfo {pages} {1707}}\BibitemShut {NoStop}%
\bibitem [{\citenamefont {Box}\ and\ \citenamefont
  {Tiao}(2011)}]{box2011bayesian}%
  \BibitemOpen
  \bibfield  {author} {\bibinfo {author} {\bibnamefont {Box}, \bibfnamefont
  {G.~E.}}, and\ \bibinfo {author} {\bibfnamefont {G.~C.}\ \bibnamefont
  {Tiao}}} (\bibinfo {year} {2011}),\ \href@noop {} {\emph {\bibinfo {title}
  {Bayesian inference in statistical analysis}}}\ (\bibinfo  {publisher} {John
  Wiley \& Sons})\BibitemShut {NoStop}%
\bibitem [{\citenamefont {Buhmann}(2000)}]{buhmann2000radial}%
  \BibitemOpen
  \bibfield  {author} {\bibinfo {author} {\bibnamefont {Buhmann}, \bibfnamefont
  {M.~D.}}} (\bibinfo {year} {2000}),\ \href@noop {} {\bibfield  {journal}
  {\bibinfo  {journal} {Acta numerica}\ }\textbf {\bibinfo {volume} {9}},\
  \bibinfo {pages} {1}}\BibitemShut {NoStop}%
\bibitem [{\citenamefont {Cabayol-Garcia}\ \emph {et~al.}(2020)\citenamefont
  {Cabayol-Garcia}, \citenamefont {Eriksen}, \citenamefont {Alarc{\'o}n},
  \citenamefont {Amara}, \citenamefont {Carretero}, \citenamefont {Casas},
  \citenamefont {Castander}, \citenamefont {Fern{\'a}ndez}, \citenamefont
  {Garc{\'\i}a-Bellido}, \citenamefont {Gaztanaga} \emph
  {et~al.}}]{cabayol2020pau}%
  \BibitemOpen
  \bibfield  {author} {\bibinfo {author} {\bibnamefont {Cabayol-Garcia},
  \bibfnamefont {L.}}, \bibinfo {author} {\bibfnamefont {M.}~\bibnamefont
  {Eriksen}}, \bibinfo {author} {\bibfnamefont {A.}~\bibnamefont
  {Alarc{\'o}n}}, \bibinfo {author} {\bibfnamefont {A.}~\bibnamefont {Amara}},
  \bibinfo {author} {\bibfnamefont {J.}~\bibnamefont {Carretero}}, \bibinfo
  {author} {\bibfnamefont {R.}~\bibnamefont {Casas}}, \bibinfo {author}
  {\bibfnamefont {F.~J.}\ \bibnamefont {Castander}}, \bibinfo {author}
  {\bibfnamefont {E.}~\bibnamefont {Fern{\'a}ndez}}, \bibinfo {author}
  {\bibfnamefont {J.}~\bibnamefont {Garc{\'\i}a-Bellido}}, \bibinfo {author}
  {\bibfnamefont {E.}~\bibnamefont {Gaztanaga}},  \emph {et~al.}} (\bibinfo
  {year} {2020}),\ \href@noop {} {\bibfield  {journal} {\bibinfo  {journal}
  {Monthly Notices of the Royal Astronomical Society}\ }\textbf {\bibinfo
  {volume} {491}}~(\bibinfo {number} {4}),\ \bibinfo {pages}
  {5392}}\BibitemShut {NoStop}%
\bibitem [{\citenamefont {Cemgil}\ \emph {et~al.}(2020)\citenamefont {Cemgil},
  \citenamefont {Ghaisas}, \citenamefont {Dvijotham}, \citenamefont {Gowal},\
  and\ \citenamefont {Kohli}}]{cemgil2020autoencoding}%
  \BibitemOpen
  \bibfield  {author} {\bibinfo {author} {\bibnamefont {Cemgil}, \bibfnamefont
  {T.}}, \bibinfo {author} {\bibfnamefont {S.}~\bibnamefont {Ghaisas}},
  \bibinfo {author} {\bibfnamefont {K.}~\bibnamefont {Dvijotham}}, \bibinfo
  {author} {\bibfnamefont {S.}~\bibnamefont {Gowal}}, and\ \bibinfo {author}
  {\bibfnamefont {P.}~\bibnamefont {Kohli}}} (\bibinfo {year} {2020}),\
  \href@noop {} {\bibfield  {journal} {\bibinfo  {journal} {Advances in Neural
  Information Processing Systems}\ }\textbf {\bibinfo {volume} {33}},\ \bibinfo
  {pages} {15077}}\BibitemShut {NoStop}%
\bibitem [{\citenamefont {Chen}\ \emph {et~al.}(2023)\citenamefont {Chen},
  \citenamefont {Wang}, \citenamefont {Zhang},\ and\ \citenamefont
  {Zhang}}]{chen2023test}%
  \BibitemOpen
  \bibfield  {author} {\bibinfo {author} {\bibnamefont {Chen}, \bibfnamefont
  {J.-F.}}, \bibinfo {author} {\bibfnamefont {Y.-C.}\ \bibnamefont {Wang}},
  \bibinfo {author} {\bibfnamefont {T.}~\bibnamefont {Zhang}}, and\ \bibinfo
  {author} {\bibfnamefont {T.-J.}\ \bibnamefont {Zhang}}} (\bibinfo {year}
  {2023}),\ \href@noop {} {\bibfield  {journal} {\bibinfo  {journal} {Physical
  Review D}\ }\textbf {\bibinfo {volume} {107}}~(\bibinfo {number} {6}),\
  \bibinfo {pages} {063517}}\BibitemShut {NoStop}%
\bibitem [{\citenamefont {Chen}\ \emph {et~al.}(2020)\citenamefont {Chen},
  \citenamefont {Dai}, \citenamefont {Liu}, \citenamefont {Chen}, \citenamefont
  {Yuan},\ and\ \citenamefont {Liu}}]{chen2020dynamic}%
  \BibitemOpen
  \bibfield  {author} {\bibinfo {author} {\bibnamefont {Chen}, \bibfnamefont
  {Y.}}, \bibinfo {author} {\bibfnamefont {X.}~\bibnamefont {Dai}}, \bibinfo
  {author} {\bibfnamefont {M.}~\bibnamefont {Liu}}, \bibinfo {author}
  {\bibfnamefont {D.}~\bibnamefont {Chen}}, \bibinfo {author} {\bibfnamefont
  {L.}~\bibnamefont {Yuan}}, and\ \bibinfo {author} {\bibfnamefont
  {Z.}~\bibnamefont {Liu}}} (\bibinfo {year} {2020}),\ in\ \href@noop {} {\emph
  {\bibinfo {booktitle} {European conference on computer vision}}}\ (\bibinfo
  {organization} {Springer})\ pp.\ \bibinfo {pages} {351--367}\BibitemShut
  {NoStop}%
\bibitem [{\citenamefont {Chib}\ and\ \citenamefont
  {Greenberg}(1995)}]{chib1995understanding}%
  \BibitemOpen
  \bibfield  {author} {\bibinfo {author} {\bibnamefont {Chib}, \bibfnamefont
  {S.}}, and\ \bibinfo {author} {\bibfnamefont {E.}~\bibnamefont {Greenberg}}}
  (\bibinfo {year} {1995}),\ \href@noop {} {\bibfield  {journal} {\bibinfo
  {journal} {The american statistician}\ }\textbf {\bibinfo {volume}
  {49}}~(\bibinfo {number} {4}),\ \bibinfo {pages} {327}}\BibitemShut {NoStop}%
\bibitem [{\citenamefont {Christensen}\ \emph {et~al.}(2001)\citenamefont
  {Christensen}, \citenamefont {Meyer}, \citenamefont {Knox},\ and\
  \citenamefont {Luey}}]{christensen2001bayesian}%
  \BibitemOpen
  \bibfield  {author} {\bibinfo {author} {\bibnamefont {Christensen},
  \bibfnamefont {N.}}, \bibinfo {author} {\bibfnamefont {R.}~\bibnamefont
  {Meyer}}, \bibinfo {author} {\bibfnamefont {L.}~\bibnamefont {Knox}}, and\
  \bibinfo {author} {\bibfnamefont {B.}~\bibnamefont {Luey}}} (\bibinfo {year}
  {2001}),\ \href@noop {} {\bibfield  {journal} {\bibinfo  {journal} {Classical
  and Quantum Gravity}\ }\textbf {\bibinfo {volume} {18}}~(\bibinfo {number}
  {14}),\ \bibinfo {pages} {2677}}\BibitemShut {NoStop}%
\bibitem [{\citenamefont {Cong}\ and\ \citenamefont
  {Zhou}(2023)}]{cong2023review}%
  \BibitemOpen
  \bibfield  {author} {\bibinfo {author} {\bibnamefont {Cong}, \bibfnamefont
  {S.}}, and\ \bibinfo {author} {\bibfnamefont {Y.}~\bibnamefont {Zhou}}}
  (\bibinfo {year} {2023}),\ \href@noop {} {\bibfield  {journal} {\bibinfo
  {journal} {Artificial Intelligence Review}\ }\textbf {\bibinfo {volume}
  {56}}~(\bibinfo {number} {3}),\ \bibinfo {pages} {1905}}\BibitemShut
  {NoStop}%
\bibitem [{\citenamefont {Cong}\ \emph {et~al.}(2014)\citenamefont {Cong},
  \citenamefont {Han}, \citenamefont {Shuo}, \citenamefont {Siqi},
  \citenamefont {Tong-Jie}, \citenamefont {Yan-Chun} \emph
  {et~al.}}]{cong2014four}%
  \BibitemOpen
  \bibfield  {author} {\bibinfo {author} {\bibnamefont {Cong}, \bibfnamefont
  {Z.}}, \bibinfo {author} {\bibfnamefont {Z.}~\bibnamefont {Han}}, \bibinfo
  {author} {\bibfnamefont {Y.}~\bibnamefont {Shuo}}, \bibinfo {author}
  {\bibfnamefont {L.}~\bibnamefont {Siqi}}, \bibinfo {author} {\bibfnamefont
  {Z.}~\bibnamefont {Tong-Jie}}, \bibinfo {author} {\bibfnamefont
  {S.}~\bibnamefont {Yan-Chun}},  \emph {et~al.}} (\bibinfo {year} {2014}),\
  \href@noop {} {\bibfield  {journal} {\bibinfo  {journal} {Research in
  Astronomy and Astrophysics}\ }\textbf {\bibinfo {volume} {14}}}\BibitemShut
  {NoStop}%
\bibitem [{\citenamefont {{Di Valentino}}\ \emph {et~al.}(2025)\citenamefont
  {{Di Valentino}}, \citenamefont {{Levi Said}}, \citenamefont {{Riess}},
  \citenamefont {{Pollo}}, \citenamefont {{Poulin}}, \citenamefont
  {{G{\'o}mez-Valent}}, \citenamefont {{Weltman}}, \citenamefont {{Palmese}},
  \citenamefont {{Huang}}, \citenamefont {{van de Bruck}}, \citenamefont
  {{Shekhar Saraf}}, \citenamefont {{Kuo}}, \citenamefont {{Uhlemann}},
  \citenamefont {{Grand{\'o}n}}, \citenamefont {{Paz}}, \citenamefont
  {{Eckert}}, \citenamefont {{Teixeira}}, \citenamefont {{Saridakis}},
  \citenamefont {{Colg{\'a}in}}, \citenamefont {{Beutler}}, \citenamefont
  {{Niedermann}}, \citenamefont {{Bajardi}}, \citenamefont {{Barenboim}},
  \citenamefont {{Gubitosi}}, \citenamefont {{Musella}}, \citenamefont
  {{Banik}}, \citenamefont {{Szapudi}}, \citenamefont {{Singal}}, \citenamefont
  {{Haro Cases}}, \citenamefont {{Chluba}}, \citenamefont {{Torrado}},
  \citenamefont {{Mifsud}}, \citenamefont {{Jedamzik}}, \citenamefont {{Said}},
  \citenamefont {{Dialektopoulos}}, \citenamefont {{Herold}}, \citenamefont
  {{Perivolaropoulos}}, \citenamefont {{Zu}}, \citenamefont {{Galbany}},
  \citenamefont {{Breuval}}, \citenamefont {{Visinelli}}, \citenamefont
  {{Escamilla}}, \citenamefont {{Anchordoqui}}, \citenamefont
  {{Sheikh-Jabbari}}, \citenamefont {{Lembo}}, \citenamefont {{Dainotti}},
  \citenamefont {{Vincenzi}}, \citenamefont {{Asgari}}, \citenamefont
  {{Gerbino}}, \citenamefont {{Forconi}}, \citenamefont {{Cantiello}},
  \citenamefont {{Moresco}}, \citenamefont {{Benetti}}, \citenamefont
  {{Sch{\"o}neberg}}, \citenamefont {{Akarsu}}, \citenamefont {{Nunes}},
  \citenamefont {{Bernardo}}, \citenamefont {{Ch{\'a}vez}}, \citenamefont
  {{Anderson}}, \citenamefont {{Watkins}}, \citenamefont {{Capozziello}},
  \citenamefont {{Li}}, \citenamefont {{Vagnozzi}}, \citenamefont {{Pan}},
  \citenamefont {{Treu}}, \citenamefont {{Irsic}}, \citenamefont {{Handley}},
  \citenamefont {{Giar{\`e}}}, \citenamefont {{Murakami}}, \citenamefont
  {{Poudou}}, \citenamefont {{Heavens}}, \citenamefont {{Kogut}}, \citenamefont
  {{Domi}}, \citenamefont {{{\L}ukasz Lenart}}, \citenamefont {{Melchiorri}},
  \citenamefont {{Vadal{\`a}}}, \citenamefont {{Amon}}, \citenamefont
  {{Bonilla}}, \citenamefont {{Reeves}}, \citenamefont {{Zhuk}}, \citenamefont
  {{Bonanno}}, \citenamefont {{{\"O}vg{\"u}n}}, \citenamefont {{Pisani}},
  \citenamefont {{Talebian}}, \citenamefont {{Abebe}}, \citenamefont
  {{Aboubrahim}}, \citenamefont {{Gonz{\'a}lez Mor{\'a}n}}, \citenamefont
  {{Kov{\'a}cs}}, \citenamefont {{Papatriantafyllou}}, \citenamefont
  {{Liddle}}, \citenamefont {{Paliathanasis}}, \citenamefont {{Borowiec}},
  \citenamefont {{Yadav}}, \citenamefont {{Yadav}}, \citenamefont {{Sen}},
  \citenamefont {{Mini Latha}}, \citenamefont {{Davis}}, \citenamefont
  {{Shajib}}, \citenamefont {{Walters}}, \citenamefont {{Idicherian Lonappan}},
  \citenamefont {{Chudaykin}}, \citenamefont {{Capodagli}}, \citenamefont {{da
  Silva}}, \citenamefont {{De Felice}}, \citenamefont {{Racioppi}},
  \citenamefont {{Soler Oficial}}, \citenamefont {{Montiel}}, \citenamefont
  {{Favale}}, \citenamefont {{Bernui}}, \citenamefont {{Velasco}},
  \citenamefont {{Heinesen}}, \citenamefont {{Bakopoulos}}, \citenamefont
  {{Chatzistavrakidis}}, \citenamefont {{Khanpour}}, \citenamefont
  {{Sathyaprakash}}, \citenamefont {{Zgirski}}, \citenamefont {{L'Huillier}},
  \citenamefont {{Famaey}}, \citenamefont {{Jain}}, \citenamefont {{Marek}},
  \citenamefont {{Zhang}}, \citenamefont {{Karmakar}}, \citenamefont
  {{Dragovich}}, \citenamefont {{Thomas}}, \citenamefont {{Correa}},
  \citenamefont {{Boiza}}, \citenamefont {{Marques}}, \citenamefont
  {{Escamilla-Rivera}}, \citenamefont {{Tzerefos}}, \citenamefont {{Zhang}},
  \citenamefont {{De Leo}}, \citenamefont {{Pfeifer}}, \citenamefont {{Lee}},
  \citenamefont {{Venter}}, \citenamefont {{Gomes}}, \citenamefont {{Roque De
  bom}}, \citenamefont {{Moreno-Pulido}}, \citenamefont {{Iosifidis}},
  \citenamefont {{Grin}}, \citenamefont {{Blixt}}, \citenamefont {{Scolnic}},
  \citenamefont {{Oriti}}, \citenamefont {{Dobrycheva}}, \citenamefont
  {{Bettoni}}, \citenamefont {{Benisty}}, \citenamefont
  {{Fern{\'a}ndez-Arenas}}, \citenamefont {{Wiltshire}}, \citenamefont
  {{Sanchez Cid}}, \citenamefont {{Tamayo}}, \citenamefont {{Valls-Gabaud}},
  \citenamefont {{Pedrotti}}, \citenamefont {{Wang}}, \citenamefont
  {{Staicova}}, \citenamefont {{Totolou}}, \citenamefont {{Rubiera-Garcia}},
  \citenamefont {{Milakovi{\'c}}}, \citenamefont {{Pesce}}, \citenamefont
  {{Sluse}}, \citenamefont {{Borka}}, \citenamefont {{Yusofi}}, \citenamefont
  {{Giusarma}}, \citenamefont {{Terlevich}}, \citenamefont {{Tomasetti}},
  \citenamefont {{Vagenas}}, \citenamefont {{Fazzari}}, \citenamefont
  {{Ferreira}}, \citenamefont {{Barakovic}}, \citenamefont
  {{Dimastrogiovanni}}, \citenamefont {{Brinch Holm}}, \citenamefont
  {{Mottola}}, \citenamefont {{{\"O}z{\"u}lker}}, \citenamefont {{Specogna}},
  \citenamefont {{Brocato}}, \citenamefont {{Jensko}}, \citenamefont
  {{Antonette Enriquez}}, \citenamefont {{Bhatia}}, \citenamefont {{Bresolin}},
  \citenamefont {{Avila}}, \citenamefont {{Bouch{\`e}}}, \citenamefont
  {{Bombacigno}}, \citenamefont {{Anagnostopoulos}}, \citenamefont {{Pace}},
  \citenamefont {{Sorrenti}}, \citenamefont {{Lobo}}, \citenamefont
  {{Courbin}}, \citenamefont {{Hansen}}, \citenamefont {{Sloan}}, \citenamefont
  {{Farrugia}}, \citenamefont {{Lynch}}, \citenamefont {{Garcia-Arroyo}},
  \citenamefont {{Raimondo}}, \citenamefont {{Lambiase}}, \citenamefont
  {{Anand}}, \citenamefont {{Poulot}}, \citenamefont {{Leon}}, \citenamefont
  {{Kouniatalis}}, \citenamefont {{Nardini}}, \citenamefont {{Cs{\"o}rnyei}},
  \citenamefont {{Galloni}},\ and\ \citenamefont
  {{Bargiacchi}}}]{2025arXiv250401669D}%
  \BibitemOpen
  \bibfield  {author} {\bibinfo {author} {\bibnamefont {{Di Valentino}},
  \bibfnamefont {E.}}, \bibinfo {author} {\bibfnamefont {J.}~\bibnamefont
  {{Levi Said}}}, \bibinfo {author} {\bibfnamefont {A.}~\bibnamefont
  {{Riess}}}, \bibinfo {author} {\bibfnamefont {A.}~\bibnamefont {{Pollo}}},
  \bibinfo {author} {\bibfnamefont {V.}~\bibnamefont {{Poulin}}}, \bibinfo
  {author} {\bibfnamefont {A.}~\bibnamefont {{G{\'o}mez-Valent}}}, \bibinfo
  {author} {\bibfnamefont {A.}~\bibnamefont {{Weltman}}}, \bibinfo {author}
  {\bibfnamefont {A.}~\bibnamefont {{Palmese}}}, \bibinfo {author}
  {\bibfnamefont {C.~D.}\ \bibnamefont {{Huang}}}, \bibinfo {author}
  {\bibfnamefont {C.}~\bibnamefont {{van de Bruck}}}, \bibinfo {author}
  {\bibfnamefont {C.}~\bibnamefont {{Shekhar Saraf}}}, \bibinfo {author}
  {\bibfnamefont {C.-Y.}\ \bibnamefont {{Kuo}}}, \bibinfo {author}
  {\bibfnamefont {C.}~\bibnamefont {{Uhlemann}}}, \bibinfo {author}
  {\bibfnamefont {D.}~\bibnamefont {{Grand{\'o}n}}}, \bibinfo {author}
  {\bibfnamefont {D.}~\bibnamefont {{Paz}}}, \bibinfo {author} {\bibfnamefont
  {D.}~\bibnamefont {{Eckert}}}, \bibinfo {author} {\bibfnamefont {E.~M.}\
  \bibnamefont {{Teixeira}}}, \bibinfo {author} {\bibfnamefont {E.~N.}\
  \bibnamefont {{Saridakis}}}, \bibinfo {author} {\bibfnamefont {E.~{\'O}.}\
  \bibnamefont {{Colg{\'a}in}}}, \bibinfo {author} {\bibfnamefont
  {F.}~\bibnamefont {{Beutler}}}, \bibinfo {author} {\bibfnamefont
  {F.}~\bibnamefont {{Niedermann}}}, \bibinfo {author} {\bibfnamefont
  {F.}~\bibnamefont {{Bajardi}}}, \bibinfo {author} {\bibfnamefont
  {G.}~\bibnamefont {{Barenboim}}}, \bibinfo {author} {\bibfnamefont
  {G.}~\bibnamefont {{Gubitosi}}}, \bibinfo {author} {\bibfnamefont
  {I.}~\bibnamefont {{Musella}}}, \bibinfo {author} {\bibfnamefont
  {I.}~\bibnamefont {{Banik}}}, \bibinfo {author} {\bibfnamefont
  {I.}~\bibnamefont {{Szapudi}}}, \bibinfo {author} {\bibfnamefont
  {J.}~\bibnamefont {{Singal}}}, \bibinfo {author} {\bibfnamefont
  {J.}~\bibnamefont {{Haro Cases}}}, \bibinfo {author} {\bibfnamefont
  {J.}~\bibnamefont {{Chluba}}}, \bibinfo {author} {\bibfnamefont
  {J.}~\bibnamefont {{Torrado}}}, \bibinfo {author} {\bibfnamefont
  {J.}~\bibnamefont {{Mifsud}}}, \bibinfo {author} {\bibfnamefont
  {K.}~\bibnamefont {{Jedamzik}}}, \bibinfo {author} {\bibfnamefont
  {K.}~\bibnamefont {{Said}}}, \bibinfo {author} {\bibfnamefont
  {K.}~\bibnamefont {{Dialektopoulos}}}, \bibinfo {author} {\bibfnamefont
  {L.}~\bibnamefont {{Herold}}}, \bibinfo {author} {\bibfnamefont
  {L.}~\bibnamefont {{Perivolaropoulos}}}, \bibinfo {author} {\bibfnamefont
  {L.}~\bibnamefont {{Zu}}}, \bibinfo {author} {\bibfnamefont {L.}~\bibnamefont
  {{Galbany}}}, \bibinfo {author} {\bibfnamefont {L.}~\bibnamefont
  {{Breuval}}}, \bibinfo {author} {\bibfnamefont {L.}~\bibnamefont
  {{Visinelli}}}, \bibinfo {author} {\bibfnamefont {L.~A.}\ \bibnamefont
  {{Escamilla}}}, \bibinfo {author} {\bibfnamefont {L.~A.}\ \bibnamefont
  {{Anchordoqui}}}, \bibinfo {author} {\bibfnamefont {M.~M.}\ \bibnamefont
  {{Sheikh-Jabbari}}}, \bibinfo {author} {\bibfnamefont {M.}~\bibnamefont
  {{Lembo}}}, \bibinfo {author} {\bibfnamefont {M.~G.}\ \bibnamefont
  {{Dainotti}}}, \bibinfo {author} {\bibfnamefont {M.}~\bibnamefont
  {{Vincenzi}}}, \bibinfo {author} {\bibfnamefont {M.}~\bibnamefont
  {{Asgari}}}, \bibinfo {author} {\bibfnamefont {M.}~\bibnamefont {{Gerbino}}},
  \bibinfo {author} {\bibfnamefont {M.}~\bibnamefont {{Forconi}}}, \bibinfo
  {author} {\bibfnamefont {M.}~\bibnamefont {{Cantiello}}}, \bibinfo {author}
  {\bibfnamefont {M.}~\bibnamefont {{Moresco}}}, \bibinfo {author}
  {\bibfnamefont {M.}~\bibnamefont {{Benetti}}}, \bibinfo {author}
  {\bibfnamefont {N.}~\bibnamefont {{Sch{\"o}neberg}}}, \bibinfo {author}
  {\bibfnamefont {{\"O}.}~\bibnamefont {{Akarsu}}}, \bibinfo {author}
  {\bibfnamefont {R.~C.}\ \bibnamefont {{Nunes}}}, \bibinfo {author}
  {\bibfnamefont {R.~C.}\ \bibnamefont {{Bernardo}}}, \bibinfo {author}
  {\bibfnamefont {R.}~\bibnamefont {{Ch{\'a}vez}}}, \bibinfo {author}
  {\bibfnamefont {R.~I.}\ \bibnamefont {{Anderson}}}, \bibinfo {author}
  {\bibfnamefont {R.}~\bibnamefont {{Watkins}}}, \bibinfo {author}
  {\bibfnamefont {S.}~\bibnamefont {{Capozziello}}}, \bibinfo {author}
  {\bibfnamefont {S.}~\bibnamefont {{Li}}}, \bibinfo {author} {\bibfnamefont
  {S.}~\bibnamefont {{Vagnozzi}}}, \bibinfo {author} {\bibfnamefont
  {S.}~\bibnamefont {{Pan}}}, \bibinfo {author} {\bibfnamefont
  {T.}~\bibnamefont {{Treu}}}, \bibinfo {author} {\bibfnamefont
  {V.}~\bibnamefont {{Irsic}}}, \bibinfo {author} {\bibfnamefont
  {W.}~\bibnamefont {{Handley}}}, \bibinfo {author} {\bibfnamefont
  {W.}~\bibnamefont {{Giar{\`e}}}}, \bibinfo {author} {\bibfnamefont
  {Y.}~\bibnamefont {{Murakami}}}, \bibinfo {author} {\bibfnamefont
  {A.}~\bibnamefont {{Poudou}}}, \bibinfo {author} {\bibfnamefont
  {A.}~\bibnamefont {{Heavens}}}, \bibinfo {author} {\bibfnamefont
  {A.}~\bibnamefont {{Kogut}}}, \bibinfo {author} {\bibfnamefont
  {A.}~\bibnamefont {{Domi}}}, \bibinfo {author} {\bibfnamefont
  {A.}~\bibnamefont {{{\L}ukasz Lenart}}}, \bibinfo {author} {\bibfnamefont
  {A.}~\bibnamefont {{Melchiorri}}}, \bibinfo {author} {\bibfnamefont
  {A.}~\bibnamefont {{Vadal{\`a}}}}, \bibinfo {author} {\bibfnamefont
  {A.}~\bibnamefont {{Amon}}}, \bibinfo {author} {\bibfnamefont
  {A.}~\bibnamefont {{Bonilla}}}, \bibinfo {author} {\bibfnamefont
  {A.}~\bibnamefont {{Reeves}}}, \bibinfo {author} {\bibfnamefont
  {A.}~\bibnamefont {{Zhuk}}}, \bibinfo {author} {\bibfnamefont
  {A.}~\bibnamefont {{Bonanno}}}, \bibinfo {author} {\bibfnamefont
  {A.}~\bibnamefont {{{\"O}vg{\"u}n}}}, \bibinfo {author} {\bibfnamefont
  {A.}~\bibnamefont {{Pisani}}}, \bibinfo {author} {\bibfnamefont
  {A.}~\bibnamefont {{Talebian}}}, \bibinfo {author} {\bibfnamefont
  {A.}~\bibnamefont {{Abebe}}}, \bibinfo {author} {\bibfnamefont
  {A.}~\bibnamefont {{Aboubrahim}}}, \bibinfo {author} {\bibfnamefont {A.~L.}\
  \bibnamefont {{Gonz{\'a}lez Mor{\'a}n}}}, \bibinfo {author} {\bibfnamefont
  {A.}~\bibnamefont {{Kov{\'a}cs}}}, \bibinfo {author} {\bibfnamefont
  {A.}~\bibnamefont {{Papatriantafyllou}}}, \bibinfo {author} {\bibfnamefont
  {A.~R.}\ \bibnamefont {{Liddle}}}, \bibinfo {author} {\bibfnamefont
  {A.}~\bibnamefont {{Paliathanasis}}}, \bibinfo {author} {\bibfnamefont
  {A.}~\bibnamefont {{Borowiec}}}, \bibinfo {author} {\bibfnamefont {A.~K.}\
  \bibnamefont {{Yadav}}}, \bibinfo {author} {\bibfnamefont {A.}~\bibnamefont
  {{Yadav}}}, \bibinfo {author} {\bibfnamefont {A.~A.}\ \bibnamefont {{Sen}}},
  \bibinfo {author} {\bibfnamefont {A.~J.~W.}\ \bibnamefont {{Mini Latha}}},
  \bibinfo {author} {\bibfnamefont {A.~C.}\ \bibnamefont {{Davis}}}, \bibinfo
  {author} {\bibfnamefont {A.~J.}\ \bibnamefont {{Shajib}}}, \bibinfo {author}
  {\bibfnamefont {A.}~\bibnamefont {{Walters}}}, \bibinfo {author}
  {\bibfnamefont {A.}~\bibnamefont {{Idicherian Lonappan}}}, \bibinfo {author}
  {\bibfnamefont {A.}~\bibnamefont {{Chudaykin}}}, \bibinfo {author}
  {\bibfnamefont {A.}~\bibnamefont {{Capodagli}}}, \bibinfo {author}
  {\bibfnamefont {A.}~\bibnamefont {{da Silva}}}, \bibinfo {author}
  {\bibfnamefont {A.}~\bibnamefont {{De Felice}}}, \bibinfo {author}
  {\bibfnamefont {A.}~\bibnamefont {{Racioppi}}}, \bibinfo {author}
  {\bibfnamefont {A.}~\bibnamefont {{Soler Oficial}}}, \bibinfo {author}
  {\bibfnamefont {A.}~\bibnamefont {{Montiel}}}, \bibinfo {author}
  {\bibfnamefont {A.}~\bibnamefont {{Favale}}}, \bibinfo {author}
  {\bibfnamefont {A.}~\bibnamefont {{Bernui}}}, \bibinfo {author}
  {\bibfnamefont {A.~C.}\ \bibnamefont {{Velasco}}}, \bibinfo {author}
  {\bibfnamefont {A.}~\bibnamefont {{Heinesen}}}, \bibinfo {author}
  {\bibfnamefont {A.}~\bibnamefont {{Bakopoulos}}}, \bibinfo {author}
  {\bibfnamefont {A.}~\bibnamefont {{Chatzistavrakidis}}}, \bibinfo {author}
  {\bibfnamefont {B.}~\bibnamefont {{Khanpour}}}, \bibinfo {author}
  {\bibfnamefont {B.~S.}\ \bibnamefont {{Sathyaprakash}}}, \bibinfo {author}
  {\bibfnamefont {B.}~\bibnamefont {{Zgirski}}}, \bibinfo {author}
  {\bibfnamefont {B.}~\bibnamefont {{L'Huillier}}}, \bibinfo {author}
  {\bibfnamefont {B.}~\bibnamefont {{Famaey}}}, \bibinfo {author}
  {\bibfnamefont {B.}~\bibnamefont {{Jain}}}, \bibinfo {author} {\bibfnamefont
  {B.}~\bibnamefont {{Marek}}}, \bibinfo {author} {\bibfnamefont
  {B.}~\bibnamefont {{Zhang}}}, \bibinfo {author} {\bibfnamefont
  {B.}~\bibnamefont {{Karmakar}}}, \bibinfo {author} {\bibfnamefont
  {B.}~\bibnamefont {{Dragovich}}}, \bibinfo {author} {\bibfnamefont
  {B.}~\bibnamefont {{Thomas}}}, \bibinfo {author} {\bibfnamefont
  {C.}~\bibnamefont {{Correa}}}, \bibinfo {author} {\bibfnamefont {C.~G.}\
  \bibnamefont {{Boiza}}}, \bibinfo {author} {\bibfnamefont {C.}~\bibnamefont
  {{Marques}}}, \bibinfo {author} {\bibfnamefont {C.}~\bibnamefont
  {{Escamilla-Rivera}}}, \bibinfo {author} {\bibfnamefont {C.}~\bibnamefont
  {{Tzerefos}}}, \bibinfo {author} {\bibfnamefont {C.}~\bibnamefont {{Zhang}}},
  \bibinfo {author} {\bibfnamefont {C.}~\bibnamefont {{De Leo}}}, \bibinfo
  {author} {\bibfnamefont {C.}~\bibnamefont {{Pfeifer}}}, \bibinfo {author}
  {\bibfnamefont {C.}~\bibnamefont {{Lee}}}, \bibinfo {author} {\bibfnamefont
  {C.}~\bibnamefont {{Venter}}}, \bibinfo {author} {\bibfnamefont
  {C.}~\bibnamefont {{Gomes}}}, \bibinfo {author} {\bibfnamefont
  {C.}~\bibnamefont {{Roque De bom}}}, \bibinfo {author} {\bibfnamefont
  {C.}~\bibnamefont {{Moreno-Pulido}}}, \bibinfo {author} {\bibfnamefont
  {D.}~\bibnamefont {{Iosifidis}}}, \bibinfo {author} {\bibfnamefont
  {D.}~\bibnamefont {{Grin}}}, \bibinfo {author} {\bibfnamefont
  {D.}~\bibnamefont {{Blixt}}}, \bibinfo {author} {\bibfnamefont
  {D.}~\bibnamefont {{Scolnic}}}, \bibinfo {author} {\bibfnamefont
  {D.}~\bibnamefont {{Oriti}}}, \bibinfo {author} {\bibfnamefont
  {D.}~\bibnamefont {{Dobrycheva}}}, \bibinfo {author} {\bibfnamefont
  {D.}~\bibnamefont {{Bettoni}}}, \bibinfo {author} {\bibfnamefont
  {D.}~\bibnamefont {{Benisty}}}, \bibinfo {author} {\bibfnamefont
  {D.}~\bibnamefont {{Fern{\'a}ndez-Arenas}}}, \bibinfo {author} {\bibfnamefont
  {D.~L.}\ \bibnamefont {{Wiltshire}}}, \bibinfo {author} {\bibfnamefont
  {D.}~\bibnamefont {{Sanchez Cid}}}, \bibinfo {author} {\bibfnamefont
  {D.}~\bibnamefont {{Tamayo}}}, \bibinfo {author} {\bibfnamefont
  {D.}~\bibnamefont {{Valls-Gabaud}}}, \bibinfo {author} {\bibfnamefont
  {D.}~\bibnamefont {{Pedrotti}}}, \bibinfo {author} {\bibfnamefont
  {D.}~\bibnamefont {{Wang}}}, \bibinfo {author} {\bibfnamefont
  {D.}~\bibnamefont {{Staicova}}}, \bibinfo {author} {\bibfnamefont
  {D.}~\bibnamefont {{Totolou}}}, \bibinfo {author} {\bibfnamefont
  {D.}~\bibnamefont {{Rubiera-Garcia}}}, \bibinfo {author} {\bibfnamefont
  {D.}~\bibnamefont {{Milakovi{\'c}}}}, \bibinfo {author} {\bibfnamefont
  {D.}~\bibnamefont {{Pesce}}}, \bibinfo {author} {\bibfnamefont
  {D.}~\bibnamefont {{Sluse}}}, \bibinfo {author} {\bibfnamefont
  {D.}~\bibnamefont {{Borka}}}, \bibinfo {author} {\bibfnamefont
  {E.}~\bibnamefont {{Yusofi}}}, \bibinfo {author} {\bibfnamefont
  {E.}~\bibnamefont {{Giusarma}}}, \bibinfo {author} {\bibfnamefont
  {E.}~\bibnamefont {{Terlevich}}}, \bibinfo {author} {\bibfnamefont
  {E.}~\bibnamefont {{Tomasetti}}}, \bibinfo {author} {\bibfnamefont {E.~C.}\
  \bibnamefont {{Vagenas}}}, \bibinfo {author} {\bibfnamefont {E.}~\bibnamefont
  {{Fazzari}}}, \bibinfo {author} {\bibfnamefont {E.~G.~M.}\ \bibnamefont
  {{Ferreira}}}, \bibinfo {author} {\bibfnamefont {E.}~\bibnamefont
  {{Barakovic}}}, \bibinfo {author} {\bibfnamefont {E.}~\bibnamefont
  {{Dimastrogiovanni}}}, \bibinfo {author} {\bibfnamefont {E.}~\bibnamefont
  {{Brinch Holm}}}, \bibinfo {author} {\bibfnamefont {E.}~\bibnamefont
  {{Mottola}}}, \bibinfo {author} {\bibfnamefont {E.}~\bibnamefont
  {{{\"O}z{\"u}lker}}}, \bibinfo {author} {\bibfnamefont {E.}~\bibnamefont
  {{Specogna}}}, \bibinfo {author} {\bibfnamefont {E.}~\bibnamefont
  {{Brocato}}}, \bibinfo {author} {\bibfnamefont {E.}~\bibnamefont {{Jensko}}},
  \bibinfo {author} {\bibfnamefont {E.}~\bibnamefont {{Antonette Enriquez}}},
  \bibinfo {author} {\bibfnamefont {E.}~\bibnamefont {{Bhatia}}}, \bibinfo
  {author} {\bibfnamefont {F.}~\bibnamefont {{Bresolin}}}, \bibinfo {author}
  {\bibfnamefont {F.}~\bibnamefont {{Avila}}}, \bibinfo {author} {\bibfnamefont
  {F.}~\bibnamefont {{Bouch{\`e}}}}, \bibinfo {author} {\bibfnamefont
  {F.}~\bibnamefont {{Bombacigno}}}, \bibinfo {author} {\bibfnamefont {F.~K.}\
  \bibnamefont {{Anagnostopoulos}}}, \bibinfo {author} {\bibfnamefont
  {F.}~\bibnamefont {{Pace}}}, \bibinfo {author} {\bibfnamefont
  {F.}~\bibnamefont {{Sorrenti}}}, \bibinfo {author} {\bibfnamefont {F.~S.~N.}\
  \bibnamefont {{Lobo}}}, \bibinfo {author} {\bibfnamefont {F.}~\bibnamefont
  {{Courbin}}}, \bibinfo {author} {\bibfnamefont {F.~K.}\ \bibnamefont
  {{Hansen}}}, \bibinfo {author} {\bibfnamefont {G.}~\bibnamefont {{Sloan}}},
  \bibinfo {author} {\bibfnamefont {G.}~\bibnamefont {{Farrugia}}}, \bibinfo
  {author} {\bibfnamefont {G.}~\bibnamefont {{Lynch}}}, \bibinfo {author}
  {\bibfnamefont {G.}~\bibnamefont {{Garcia-Arroyo}}}, \bibinfo {author}
  {\bibfnamefont {G.}~\bibnamefont {{Raimondo}}}, \bibinfo {author}
  {\bibfnamefont {G.}~\bibnamefont {{Lambiase}}}, \bibinfo {author}
  {\bibfnamefont {G.~S.}\ \bibnamefont {{Anand}}}, \bibinfo {author}
  {\bibfnamefont {G.}~\bibnamefont {{Poulot}}}, \bibinfo {author}
  {\bibfnamefont {G.}~\bibnamefont {{Leon}}}, \bibinfo {author} {\bibfnamefont
  {G.}~\bibnamefont {{Kouniatalis}}}, \bibinfo {author} {\bibfnamefont
  {G.}~\bibnamefont {{Nardini}}}, \bibinfo {author} {\bibfnamefont
  {G.}~\bibnamefont {{Cs{\"o}rnyei}}}, \bibinfo {author} {\bibfnamefont
  {G.}~\bibnamefont {{Galloni}}}, and\ \bibinfo {author} {\bibfnamefont
  {G.}~\bibnamefont {{Bargiacchi}}}} (\bibinfo {year} {2025}),\ \href
  {https://doi.org/10.48550/arXiv.2504.01669} {\bibfield  {journal} {\bibinfo
  {journal} {arXiv e-prints}\ ,\ \bibinfo {eid} {arXiv:2504.01669}}}\Eprint
  {https://arxiv.org/abs/2504.01669} {arXiv:2504.01669 [astro-ph.CO]}
  \BibitemShut {NoStop}%
\bibitem [{\citenamefont {Fan}(2002)}]{fan2002travelling}%
  \BibitemOpen
  \bibfield  {author} {\bibinfo {author} {\bibnamefont {Fan}, \bibfnamefont
  {E.}}} (\bibinfo {year} {2002}),\ \href@noop {} {\bibfield  {journal}
  {\bibinfo  {journal} {Physics Letters A}\ }\textbf {\bibinfo {volume}
  {300}}~(\bibinfo {number} {2-3}),\ \bibinfo {pages} {243}}\BibitemShut
  {NoStop}%
\bibitem [{\citenamefont {Foreman-Mackey}\ \emph {et~al.}(2013)\citenamefont
  {Foreman-Mackey}, \citenamefont {Hogg}, \citenamefont {Lang},\ and\
  \citenamefont {Goodman}}]{foreman2013emcee}%
  \BibitemOpen
  \bibfield  {author} {\bibinfo {author} {\bibnamefont {Foreman-Mackey},
  \bibfnamefont {D.}}, \bibinfo {author} {\bibfnamefont {D.~W.}\ \bibnamefont
  {Hogg}}, \bibinfo {author} {\bibfnamefont {D.}~\bibnamefont {Lang}}, and\
  \bibinfo {author} {\bibfnamefont {J.}~\bibnamefont {Goodman}}} (\bibinfo
  {year} {2013}),\ \href@noop {} {\bibfield  {journal} {\bibinfo  {journal}
  {Publications of the Astronomical Society of the Pacific}\ }\textbf {\bibinfo
  {volume} {125}}~(\bibinfo {number} {925}),\ \bibinfo {pages}
  {306}}\BibitemShut {NoStop}%
\bibitem [{\citenamefont {Freedman}\ and\ \citenamefont
  {Madore}(2010)}]{freedman2010hubble}%
  \BibitemOpen
  \bibfield  {author} {\bibinfo {author} {\bibnamefont {Freedman},
  \bibfnamefont {W.~L.}}, and\ \bibinfo {author} {\bibfnamefont {B.~F.}\
  \bibnamefont {Madore}}} (\bibinfo {year} {2010}),\ \href@noop {} {\bibfield
  {journal} {\bibinfo  {journal} {Annual Review of Astronomy and Astrophysics}\
  }\textbf {\bibinfo {volume} {48}}~(\bibinfo {number} {1}),\ \bibinfo {pages}
  {673}}\BibitemShut {NoStop}%
\bibitem [{\citenamefont {Friedman}(1922)}]{friedman1922krummung}%
  \BibitemOpen
  \bibfield  {author} {\bibinfo {author} {\bibnamefont {Friedman},
  \bibfnamefont {A.}}} (\bibinfo {year} {1922}),\ \href@noop {} {\bibfield
  {journal} {\bibinfo  {journal} {Zeitschrift f{\"u}r Physik}\ }\textbf
  {\bibinfo {volume} {10}}~(\bibinfo {number} {1}),\ \bibinfo {pages}
  {377}}\BibitemShut {NoStop}%
\bibitem [{\citenamefont {Goodfellow}\ \emph {et~al.}(2014)\citenamefont
  {Goodfellow}, \citenamefont {Pouget-Abadie}, \citenamefont {Mirza},
  \citenamefont {Xu}, \citenamefont {Warde-Farley}, \citenamefont {Ozair},
  \citenamefont {Courville},\ and\ \citenamefont
  {Bengio}}]{goodfellow2014generative}%
  \BibitemOpen
  \bibfield  {author} {\bibinfo {author} {\bibnamefont {Goodfellow},
  \bibfnamefont {I.}}, \bibinfo {author} {\bibfnamefont {J.}~\bibnamefont
  {Pouget-Abadie}}, \bibinfo {author} {\bibfnamefont {M.}~\bibnamefont
  {Mirza}}, \bibinfo {author} {\bibfnamefont {B.}~\bibnamefont {Xu}}, \bibinfo
  {author} {\bibfnamefont {D.}~\bibnamefont {Warde-Farley}}, \bibinfo {author}
  {\bibfnamefont {S.}~\bibnamefont {Ozair}}, \bibinfo {author} {\bibfnamefont
  {A.}~\bibnamefont {Courville}}, and\ \bibinfo {author} {\bibfnamefont
  {Y.}~\bibnamefont {Bengio}}} (\bibinfo {year} {2014}),\ \href@noop {}
  {\bibfield  {journal} {\bibinfo  {journal} {Advances in neural information
  processing systems}\ }\textbf {\bibinfo {volume} {27}}}\BibitemShut {NoStop}%
\bibitem [{\citenamefont {Goodman}\ and\ \citenamefont
  {Weare}(2010)}]{goodman2010ensemble}%
  \BibitemOpen
  \bibfield  {author} {\bibinfo {author} {\bibnamefont {Goodman}, \bibfnamefont
  {J.}}, and\ \bibinfo {author} {\bibfnamefont {J.}~\bibnamefont {Weare}}}
  (\bibinfo {year} {2010}),\ \href@noop {} {\bibfield  {journal} {\bibinfo
  {journal} {Communications in applied mathematics and computational science}\
  }\textbf {\bibinfo {volume} {5}}~(\bibinfo {number} {1}),\ \bibinfo {pages}
  {65}}\BibitemShut {NoStop}%
\bibitem [{\citenamefont {Gómez-Vargas}\ \emph
  {et~al.}(2023{\natexlab{a}})\citenamefont {Gómez-Vargas}, \citenamefont
  {Briones~Andrade},\ and\ \citenamefont {Vázquez}}]{G_mez_Vargas_2023}%
  \BibitemOpen
  \bibfield  {author} {\bibinfo {author} {\bibnamefont {Gómez-Vargas},
  \bibfnamefont {I.}}, \bibinfo {author} {\bibfnamefont {J.}~\bibnamefont
  {Briones~Andrade}}, and\ \bibinfo {author} {\bibfnamefont {J.~A.}\
  \bibnamefont {Vázquez}}} (\bibinfo {year} {2023}{\natexlab{a}}),\ \href
  {https://doi.org/10.1103/physrevd.107.043509} {\bibfield  {journal} {\bibinfo
   {journal} {Physical Review D}\ }\textbf {\bibinfo {volume} {107}}~(\bibinfo
  {number} {4}),\ 10.1103/physrevd.107.043509}\BibitemShut {NoStop}%
\bibitem [{\citenamefont {Gómez-Vargas}\ \emph
  {et~al.}(2023{\natexlab{b}})\citenamefont {Gómez-Vargas}, \citenamefont
  {Medel-Esquivel}, \citenamefont {García-Salcedo},\ and\ \citenamefont
  {Vázquez}}]{G_mez_Vargas1_2023}%
  \BibitemOpen
  \bibfield  {author} {\bibinfo {author} {\bibnamefont {Gómez-Vargas},
  \bibfnamefont {I.}}, \bibinfo {author} {\bibfnamefont {R.}~\bibnamefont
  {Medel-Esquivel}}, \bibinfo {author} {\bibfnamefont {R.}~\bibnamefont
  {García-Salcedo}}, and\ \bibinfo {author} {\bibfnamefont {J.~A.}\
  \bibnamefont {Vázquez}}} (\bibinfo {year} {2023}{\natexlab{b}}),\ \href
  {https://doi.org/10.1140/epjc/s10052-023-11435-9} {\bibfield  {journal}
  {\bibinfo  {journal} {The European Physical Journal C}\ }\textbf {\bibinfo
  {volume} {83}}~(\bibinfo {number} {4}),\
  10.1140/epjc/s10052-023-11435-9}\BibitemShut {NoStop}%
\bibitem [{\citenamefont {He}\ \emph {et~al.}(2019)\citenamefont {He},
  \citenamefont {Liu},\ and\ \citenamefont {Tao}}]{he2019control}%
  \BibitemOpen
  \bibfield  {author} {\bibinfo {author} {\bibnamefont {He}, \bibfnamefont
  {F.}}, \bibinfo {author} {\bibfnamefont {T.}~\bibnamefont {Liu}}, and\
  \bibinfo {author} {\bibfnamefont {D.}~\bibnamefont {Tao}}} (\bibinfo {year}
  {2019}),\ \href@noop {} {\bibfield  {journal} {\bibinfo  {journal} {Advances
  in neural information processing systems}\ }\textbf {\bibinfo {volume}
  {32}}}\BibitemShut {NoStop}%
\bibitem [{\citenamefont {Hecht-Nielsen}(1992)}]{hecht1992theory}%
  \BibitemOpen
  \bibfield  {author} {\bibinfo {author} {\bibnamefont {Hecht-Nielsen},
  \bibfnamefont {R.}}} (\bibinfo {year} {1992}),\ in\ \href@noop {} {\emph
  {\bibinfo {booktitle} {Neural networks for perception}}}\ (\bibinfo
  {publisher} {Elsevier})\ pp.\ \bibinfo {pages} {65--93}\BibitemShut {NoStop}%
\bibitem [{\citenamefont {Jackson}(2015)}]{jackson2015hubble}%
  \BibitemOpen
  \bibfield  {author} {\bibinfo {author} {\bibnamefont {Jackson}, \bibfnamefont
  {N.}}} (\bibinfo {year} {2015}),\ \href@noop {} {\bibfield  {journal}
  {\bibinfo  {journal} {Living reviews in relativity}\ }\textbf {\bibinfo
  {volume} {18}}~(\bibinfo {number} {1}),\ \bibinfo {pages} {2}}\BibitemShut
  {NoStop}%
\bibitem [{\citenamefont {Jesus}\ \emph {et~al.}(2017)\citenamefont {Jesus},
  \citenamefont {Gregório}, \citenamefont {Andrade-Oliveira}, \citenamefont
  {Valentim},\ and\ \citenamefont {Matos}}]{2017Bayesian}%
  \BibitemOpen
  \bibfield  {author} {\bibinfo {author} {\bibnamefont {Jesus}, \bibfnamefont
  {J.~F.}}, \bibinfo {author} {\bibfnamefont {T.}~\bibnamefont {Gregório}},
  \bibinfo {author} {\bibfnamefont {F.}~\bibnamefont {Andrade-Oliveira}},
  \bibinfo {author} {\bibfnamefont {R.}~\bibnamefont {Valentim}}, and\ \bibinfo
  {author} {\bibfnamefont {C.}~\bibnamefont {Matos}}} (\bibinfo {year}
  {2017}),\ \href@noop {} {\bibfield  {journal} {\bibinfo  {journal} {Monthly
  Notices of the Royal Astronomical Society}\ }~(\bibinfo {number} {3}),\
  \bibinfo {pages} {3}}\BibitemShut {NoStop}%
\bibitem [{\citenamefont {Jiao}\ \emph {et~al.}(2023)\citenamefont {Jiao},
  \citenamefont {Borghi}, \citenamefont {Moresco},\ and\ \citenamefont
  {Zhang}}]{jiao2023new}%
  \BibitemOpen
  \bibfield  {author} {\bibinfo {author} {\bibnamefont {Jiao}, \bibfnamefont
  {K.}}, \bibinfo {author} {\bibfnamefont {N.}~\bibnamefont {Borghi}}, \bibinfo
  {author} {\bibfnamefont {M.}~\bibnamefont {Moresco}}, and\ \bibinfo {author}
  {\bibfnamefont {T.-J.}\ \bibnamefont {Zhang}}} (\bibinfo {year} {2023}),\
  \href@noop {} {\bibfield  {journal} {\bibinfo  {journal} {The Astrophysical
  Journal Supplement Series}\ }\textbf {\bibinfo {volume} {265}}~(\bibinfo
  {number} {2}),\ \bibinfo {pages} {48}}\BibitemShut {NoStop}%
\bibitem [{\citenamefont {Jimenez}\ \emph
  {et~al.}(2003{\natexlab{a}})\citenamefont {Jimenez}, \citenamefont {Verde},
  \citenamefont {Treu},\ and\ \citenamefont {Stern}}]{jimenez2003constraints}%
  \BibitemOpen
  \bibfield  {author} {\bibinfo {author} {\bibnamefont {Jimenez}, \bibfnamefont
  {R.}}, \bibinfo {author} {\bibfnamefont {L.}~\bibnamefont {Verde}}, \bibinfo
  {author} {\bibfnamefont {T.}~\bibnamefont {Treu}}, and\ \bibinfo {author}
  {\bibfnamefont {D.}~\bibnamefont {Stern}}} (\bibinfo {year}
  {2003}{\natexlab{a}}),\ \href@noop {} {\bibfield  {journal} {\bibinfo
  {journal} {The Astrophysical Journal}\ }\textbf {\bibinfo {volume}
  {593}}~(\bibinfo {number} {2}),\ \bibinfo {pages} {622}}\BibitemShut
  {NoStop}%
\bibitem [{\citenamefont {Jimenez}\ \emph
  {et~al.}(2003{\natexlab{b}})\citenamefont {Jimenez}, \citenamefont {Verde},
  \citenamefont {Treu},\ and\ \citenamefont {Stern}}]{Jimenez_2003}%
  \BibitemOpen
  \bibfield  {author} {\bibinfo {author} {\bibnamefont {Jimenez}, \bibfnamefont
  {R.}}, \bibinfo {author} {\bibfnamefont {L.}~\bibnamefont {Verde}}, \bibinfo
  {author} {\bibfnamefont {T.}~\bibnamefont {Treu}}, and\ \bibinfo {author}
  {\bibfnamefont {D.}~\bibnamefont {Stern}}} (\bibinfo {year}
  {2003}{\natexlab{b}}),\ \href {https://doi.org/10.1086/376595} {\bibfield
  {journal} {\bibinfo  {journal} {The Astrophysical Journal}\ }\textbf
  {\bibinfo {volume} {593}}~(\bibinfo {number} {2}),\ \bibinfo {pages}
  {622}}\BibitemShut {NoStop}%
\bibitem [{\citenamefont {Kenton}\ and\ \citenamefont
  {Toutanova}(2019)}]{kenton2019bert}%
  \BibitemOpen
  \bibfield  {author} {\bibinfo {author} {\bibnamefont {Kenton}, \bibfnamefont
  {J.~D. M.-W.~C.}}, and\ \bibinfo {author} {\bibfnamefont {L.~K.}\
  \bibnamefont {Toutanova}}} (\bibinfo {year} {2019}),\ in\ \href@noop {}
  {\emph {\bibinfo {booktitle} {Proceedings of naacL-HLT}}},\ Vol.~\bibinfo
  {volume} {1}\ (\bibinfo {organization} {Minneapolis, Minnesota})\BibitemShut
  {NoStop}%
\bibitem [{\citenamefont {Kyurkchiev}\ and\ \citenamefont
  {Markov}(2015)}]{kyurkchiev2015sigmoid}%
  \BibitemOpen
  \bibfield  {author} {\bibinfo {author} {\bibnamefont {Kyurkchiev},
  \bibfnamefont {N.}}, and\ \bibinfo {author} {\bibfnamefont {S.}~\bibnamefont
  {Markov}}} (\bibinfo {year} {2015}),\ \href@noop {} {\bibfield  {journal}
  {\bibinfo  {journal} {LAP LAMBERT Academic Publishing, Saarbrucken}\ }\textbf
  {\bibinfo {volume} {4}}}\BibitemShut {NoStop}%
\bibitem [{\citenamefont {Lewis}\ and\ \citenamefont
  {Bridle}(2002)}]{lewis2002cosmological}%
  \BibitemOpen
  \bibfield  {author} {\bibinfo {author} {\bibnamefont {Lewis}, \bibfnamefont
  {A.}}, and\ \bibinfo {author} {\bibfnamefont {S.}~\bibnamefont {Bridle}}}
  (\bibinfo {year} {2002}),\ \href@noop {} {\bibfield  {journal} {\bibinfo
  {journal} {Physical Review D}\ }\textbf {\bibinfo {volume} {66}}~(\bibinfo
  {number} {10}),\ \bibinfo {pages} {103511}}\BibitemShut {NoStop}%
\bibitem [{\citenamefont {Lu}\ \emph {et~al.}(2022{\natexlab{a}})\citenamefont
  {Lu}, \citenamefont {Jiao}, \citenamefont {Zhang}, \citenamefont {Zhang},\
  and\ \citenamefont {Zhu}}]{lu2022toward}%
  \BibitemOpen
  \bibfield  {author} {\bibinfo {author} {\bibnamefont {Lu}, \bibfnamefont
  {C.-Z.}}, \bibinfo {author} {\bibfnamefont {K.}~\bibnamefont {Jiao}},
  \bibinfo {author} {\bibfnamefont {T.}~\bibnamefont {Zhang}}, \bibinfo
  {author} {\bibfnamefont {T.-J.}\ \bibnamefont {Zhang}}, and\ \bibinfo
  {author} {\bibfnamefont {M.}~\bibnamefont {Zhu}}} (\bibinfo {year}
  {2022}{\natexlab{a}}),\ \href@noop {} {\bibfield  {journal} {\bibinfo
  {journal} {Physics of the Dark Universe}\ }\textbf {\bibinfo {volume} {37}},\
  \bibinfo {pages} {101088}}\BibitemShut {NoStop}%
\bibitem [{\citenamefont {Lu}\ \emph {et~al.}(2022{\natexlab{b}})\citenamefont
  {Lu}, \citenamefont {Zhang},\ and\ \citenamefont
  {Zhang}}]{lu2022statistical}%
  \BibitemOpen
  \bibfield  {author} {\bibinfo {author} {\bibnamefont {Lu}, \bibfnamefont
  {C.-Z.}}, \bibinfo {author} {\bibfnamefont {T.}~\bibnamefont {Zhang}}, and\
  \bibinfo {author} {\bibfnamefont {T.-J.}\ \bibnamefont {Zhang}}} (\bibinfo
  {year} {2022}{\natexlab{b}}),\ \href@noop {} {\bibinfo  {journal} {arXiv
  preprint arXiv:2208.05639}\ }\BibitemShut {NoStop}%
\bibitem [{\citenamefont {Luan}\ \emph {et~al.}(2023)\citenamefont {Luan},
  \citenamefont {Tao}, \citenamefont {Zhao}, \citenamefont {Huang},
  \citenamefont {Li}, \citenamefont {Liu}, \citenamefont {Wang}, \citenamefont
  {Liu}, \citenamefont {Zhang}, \citenamefont {Gajjar} \emph
  {et~al.}}]{luan2023multibeam}%
  \BibitemOpen
\bibfield  {journal} {  }\bibfield  {author} {\bibinfo {author} {\bibnamefont
  {Luan}, \bibfnamefont {X.-H.}}, \bibinfo {author} {\bibfnamefont {Z.-Z.}\
  \bibnamefont {Tao}}, \bibinfo {author} {\bibfnamefont {H.-C.}\ \bibnamefont
  {Zhao}}, \bibinfo {author} {\bibfnamefont {B.-L.}\ \bibnamefont {Huang}},
  \bibinfo {author} {\bibfnamefont {S.-Y.}\ \bibnamefont {Li}}, \bibinfo
  {author} {\bibfnamefont {C.}~\bibnamefont {Liu}}, \bibinfo {author}
  {\bibfnamefont {H.-F.}\ \bibnamefont {Wang}}, \bibinfo {author}
  {\bibfnamefont {W.-F.}\ \bibnamefont {Liu}}, \bibinfo {author} {\bibfnamefont
  {T.-J.}\ \bibnamefont {Zhang}}, \bibinfo {author} {\bibfnamefont
  {V.}~\bibnamefont {Gajjar}},  \emph {et~al.}} (\bibinfo {year} {2023}),\
  \href@noop {} {\bibfield  {journal} {\bibinfo  {journal} {The Astronomical
  Journal}\ }\textbf {\bibinfo {volume} {165}}~(\bibinfo {number} {3}),\
  \bibinfo {pages} {132}}\BibitemShut {NoStop}%
\bibitem [{\citenamefont {Ma}\ and\ \citenamefont {Zhang}(2011)}]{ma2011power}%
  \BibitemOpen
  \bibfield  {author} {\bibinfo {author} {\bibnamefont {Ma}, \bibfnamefont
  {C.}}, and\ \bibinfo {author} {\bibfnamefont {T.-J.}\ \bibnamefont {Zhang}}}
  (\bibinfo {year} {2011}),\ \href@noop {} {\bibfield  {journal} {\bibinfo
  {journal} {The Astrophysical Journal}\ }\textbf {\bibinfo {volume}
  {730}}~(\bibinfo {number} {2}),\ \bibinfo {pages} {74}}\BibitemShut {NoStop}%
\bibitem [{\citenamefont {MacKay}\ \emph {et~al.}(1998)\citenamefont {MacKay}
  \emph {et~al.}}]{mackay1998introduction}%
  \BibitemOpen
  \bibfield  {author} {\bibinfo {author} {\bibnamefont {MacKay}, \bibfnamefont
  {D.~J.}},  \emph {et~al.}} (\bibinfo {year} {1998}),\ \href@noop {}
  {\bibfield  {journal} {\bibinfo  {journal} {NATO ASI series F computer and
  systems sciences}\ }\textbf {\bibinfo {volume} {168}},\ \bibinfo {pages}
  {133}}\BibitemShut {NoStop}%
\bibitem [{\citenamefont {Moresco}(2015)}]{10.1093/mnrasl/slv037}%
  \BibitemOpen
  \bibfield  {author} {\bibinfo {author} {\bibnamefont {Moresco}, \bibfnamefont
  {M.}}} (\bibinfo {year} {2015}),\ \href
  {https://doi.org/10.1093/mnrasl/slv037} {\bibfield  {journal} {\bibinfo
  {journal} {Monthly Notices of the Royal Astronomical Society: Letters}\
  }\textbf {\bibinfo {volume} {450}}~(\bibinfo {number} {1}),\ \bibinfo {pages}
  {L16}},\ \Eprint
  {https://arxiv.org/abs/https://academic.oup.com/mnrasl/article-pdf/450/1/L16/3083577/slv037.pdf}
  {https://academic.oup.com/mnrasl/article-pdf/450/1/L16/3083577/slv037.pdf}
  \BibitemShut {NoStop}%
\bibitem [{\citenamefont
  {Moresco}(2024)}]{moresco2024measuringexpansionhistoryuniverse}%
  \BibitemOpen
  \bibfield  {author} {\bibinfo {author} {\bibnamefont {Moresco}, \bibfnamefont
  {M.}}} (\bibinfo {year} {2024}),\ \href {https://arxiv.org/abs/2412.01994}
  {\enquote {\bibinfo {title} {Measuring the expansion history of the universe
  with cosmic chronometers},}\ }\Eprint {https://arxiv.org/abs/2412.01994}
  {arXiv:2412.01994 [astro-ph.CO]} \BibitemShut {NoStop}%
\bibitem [{\citenamefont {Moresco}\ \emph
  {et~al.}(2012{\natexlab{a}})\citenamefont {Moresco}, \citenamefont {Cimatti},
  \citenamefont {Jimenez}, \citenamefont {Pozzetti}, \citenamefont {Zamorani},
  \citenamefont {Bolzonella}, \citenamefont {Dunlop}, \citenamefont
  {Lamareille}, \citenamefont {Mignoli}, \citenamefont {Pearce} \emph
  {et~al.}}]{moresco2012improved}%
  \BibitemOpen
  \bibfield  {author} {\bibinfo {author} {\bibnamefont {Moresco}, \bibfnamefont
  {M.}}, \bibinfo {author} {\bibfnamefont {A.}~\bibnamefont {Cimatti}},
  \bibinfo {author} {\bibfnamefont {R.}~\bibnamefont {Jimenez}}, \bibinfo
  {author} {\bibfnamefont {L.}~\bibnamefont {Pozzetti}}, \bibinfo {author}
  {\bibfnamefont {G.}~\bibnamefont {Zamorani}}, \bibinfo {author}
  {\bibfnamefont {M.}~\bibnamefont {Bolzonella}}, \bibinfo {author}
  {\bibfnamefont {J.}~\bibnamefont {Dunlop}}, \bibinfo {author} {\bibfnamefont
  {F.}~\bibnamefont {Lamareille}}, \bibinfo {author} {\bibfnamefont
  {M.}~\bibnamefont {Mignoli}}, \bibinfo {author} {\bibfnamefont
  {H.}~\bibnamefont {Pearce}},  \emph {et~al.}} (\bibinfo {year}
  {2012}{\natexlab{a}}),\ \href@noop {} {\bibfield  {journal} {\bibinfo
  {journal} {Journal of Cosmology and Astroparticle Physics}\ }\textbf
  {\bibinfo {volume} {2012}}~(\bibinfo {number} {08}),\ \bibinfo {pages}
  {006}}\BibitemShut {NoStop}%
\bibitem [{\citenamefont {Moresco}\ \emph {et~al.}(2020)\citenamefont
  {Moresco}, \citenamefont {Jimenez}, \citenamefont {Verde}, \citenamefont
  {Cimatti},\ and\ \citenamefont {Pozzetti}}]{moresco2020setting}%
  \BibitemOpen
  \bibfield  {author} {\bibinfo {author} {\bibnamefont {Moresco}, \bibfnamefont
  {M.}}, \bibinfo {author} {\bibfnamefont {R.}~\bibnamefont {Jimenez}},
  \bibinfo {author} {\bibfnamefont {L.}~\bibnamefont {Verde}}, \bibinfo
  {author} {\bibfnamefont {A.}~\bibnamefont {Cimatti}}, and\ \bibinfo {author}
  {\bibfnamefont {L.}~\bibnamefont {Pozzetti}}} (\bibinfo {year} {2020}),\
  \href@noop {} {\bibfield  {journal} {\bibinfo  {journal} {The Astrophysical
  Journal}\ }\textbf {\bibinfo {volume} {898}}~(\bibinfo {number} {1}),\
  \bibinfo {pages} {82}}\BibitemShut {NoStop}%
\bibitem [{\citenamefont {Moresco}\ \emph {et~al.}(2018)\citenamefont
  {Moresco}, \citenamefont {Jimenez}, \citenamefont {Verde}, \citenamefont
  {Pozzetti}, \citenamefont {Cimatti},\ and\ \citenamefont
  {Citro}}]{moresco2018setting}%
  \BibitemOpen
  \bibfield  {author} {\bibinfo {author} {\bibnamefont {Moresco}, \bibfnamefont
  {M.}}, \bibinfo {author} {\bibfnamefont {R.}~\bibnamefont {Jimenez}},
  \bibinfo {author} {\bibfnamefont {L.}~\bibnamefont {Verde}}, \bibinfo
  {author} {\bibfnamefont {L.}~\bibnamefont {Pozzetti}}, \bibinfo {author}
  {\bibfnamefont {A.}~\bibnamefont {Cimatti}}, and\ \bibinfo {author}
  {\bibfnamefont {A.}~\bibnamefont {Citro}}} (\bibinfo {year} {2018}),\
  \href@noop {} {\bibfield  {journal} {\bibinfo  {journal} {The Astrophysical
  Journal}\ }\textbf {\bibinfo {volume} {868}}~(\bibinfo {number} {2}),\
  \bibinfo {pages} {84}}\BibitemShut {NoStop}%
\bibitem [{\citenamefont {Moresco}\ \emph {et~al.}(2016)\citenamefont
  {Moresco}, \citenamefont {Pozzetti}, \citenamefont {Cimatti}, \citenamefont
  {Jimenez}, \citenamefont {Maraston}, \citenamefont {Verde}, \citenamefont
  {Thomas}, \citenamefont {Citro}, \citenamefont {Tojeiro},\ and\ \citenamefont
  {Wilkinson}}]{moresco20166}%
  \BibitemOpen
  \bibfield  {author} {\bibinfo {author} {\bibnamefont {Moresco}, \bibfnamefont
  {M.}}, \bibinfo {author} {\bibfnamefont {L.}~\bibnamefont {Pozzetti}},
  \bibinfo {author} {\bibfnamefont {A.}~\bibnamefont {Cimatti}}, \bibinfo
  {author} {\bibfnamefont {R.}~\bibnamefont {Jimenez}}, \bibinfo {author}
  {\bibfnamefont {C.}~\bibnamefont {Maraston}}, \bibinfo {author}
  {\bibfnamefont {L.}~\bibnamefont {Verde}}, \bibinfo {author} {\bibfnamefont
  {D.}~\bibnamefont {Thomas}}, \bibinfo {author} {\bibfnamefont
  {A.}~\bibnamefont {Citro}}, \bibinfo {author} {\bibfnamefont
  {R.}~\bibnamefont {Tojeiro}}, and\ \bibinfo {author} {\bibfnamefont
  {D.}~\bibnamefont {Wilkinson}}} (\bibinfo {year} {2016}),\ \href@noop {}
  {\bibfield  {journal} {\bibinfo  {journal} {Journal of Cosmology and
  Astroparticle Physics}\ }\textbf {\bibinfo {volume} {2016}}~(\bibinfo
  {number} {05}),\ \bibinfo {pages} {014}}\BibitemShut {NoStop}%
\bibitem [{\citenamefont {Moresco}\ \emph
  {et~al.}(2012{\natexlab{b}})\citenamefont {Moresco}, \citenamefont {Verde},
  \citenamefont {Pozzetti}, \citenamefont {Jimenez},\ and\ \citenamefont
  {Cimatti}}]{2012New}%
  \BibitemOpen
  \bibfield  {author} {\bibinfo {author} {\bibnamefont {Moresco}, \bibfnamefont
  {M.}}, \bibinfo {author} {\bibfnamefont {L.}~\bibnamefont {Verde}}, \bibinfo
  {author} {\bibfnamefont {L.}~\bibnamefont {Pozzetti}}, \bibinfo {author}
  {\bibfnamefont {R.}~\bibnamefont {Jimenez}}, and\ \bibinfo {author}
  {\bibfnamefont {A.}~\bibnamefont {Cimatti}}} (\bibinfo {year}
  {2012}{\natexlab{b}}),\ \href@noop {} {\bibfield  {journal} {\bibinfo
  {journal} {Journal of Cosmology and Astroparticle Physics}\ }\textbf
  {\bibinfo {volume} {2012}}~(\bibinfo {number} {07}),\ \bibinfo {pages}
  {053}}\BibitemShut {NoStop}%
\bibitem [{\citenamefont {Mukesh}\ \emph {et~al.}(2022)\citenamefont {Mukesh},
  \citenamefont {Ippatapu~Venkata}, \citenamefont {Chereddy}, \citenamefont
  {Anbazhagan},\ and\ \citenamefont {Oviya}}]{mukesh2022variational}%
  \BibitemOpen
  \bibfield  {author} {\bibinfo {author} {\bibnamefont {Mukesh}, \bibfnamefont
  {K.}}, \bibinfo {author} {\bibfnamefont {S.}~\bibnamefont
  {Ippatapu~Venkata}}, \bibinfo {author} {\bibfnamefont {S.}~\bibnamefont
  {Chereddy}}, \bibinfo {author} {\bibfnamefont {E.}~\bibnamefont
  {Anbazhagan}}, and\ \bibinfo {author} {\bibfnamefont {I.}~\bibnamefont
  {Oviya}}} (\bibinfo {year} {2022}),\ in\ \href@noop {} {\emph {\bibinfo
  {booktitle} {International Conference on Innovative Computing and
  Communications: Proceedings of ICICC 2022, Volume 1}}}\ (\bibinfo
  {organization} {Springer})\ pp.\ \bibinfo {pages} {761--768}\BibitemShut
  {NoStop}%
\bibitem [{\citenamefont {Niu}\ and\ \citenamefont
  {Zhang}(2023)}]{niu2023cosmological}%
  \BibitemOpen
  \bibfield  {author} {\bibinfo {author} {\bibnamefont {Niu}, \bibfnamefont
  {J.}}, and\ \bibinfo {author} {\bibfnamefont {T.-J.}\ \bibnamefont {Zhang}}}
  (\bibinfo {year} {2023}),\ \href@noop {} {\bibfield  {journal} {\bibinfo
  {journal} {Physics of the Dark Universe}\ }\textbf {\bibinfo {volume} {39}},\
  \bibinfo {pages} {101147}}\BibitemShut {NoStop}%
\bibitem [{\citenamefont {Norris}(1998)}]{norris1998markov}%
  \BibitemOpen
  \bibfield  {author} {\bibinfo {author} {\bibnamefont {Norris}, \bibfnamefont
  {J.~R.}}} (\bibinfo {year} {1998}),\ \href@noop {} {\emph {\bibinfo {title}
  {Markov chains}}},\ \bibinfo {number} {2}\ (\bibinfo  {publisher} {Cambridge
  university press})\BibitemShut {NoStop}%
\bibitem [{\citenamefont {Oliphant}\ \emph {et~al.}(2006)\citenamefont
  {Oliphant} \emph {et~al.}}]{oliphant2006guide}%
  \BibitemOpen
  \bibfield  {author} {\bibinfo {author} {\bibnamefont {Oliphant},
  \bibfnamefont {T.~E.}},  \emph {et~al.}} (\bibinfo {year} {2006}),\
  \href@noop {} {\emph {\bibinfo {title} {Guide to numpy}}},\ Vol.~\bibinfo
  {volume} {1}\ (\bibinfo  {publisher} {Trelgol Publishing USA})\BibitemShut
  {NoStop}%
\bibitem [{\citenamefont {{Pan}}\ \emph {et~al.}(2020)\citenamefont {{Pan}},
  \citenamefont {{Liu}}, \citenamefont {{Forero-Romero}}, \citenamefont
  {{Sabiu}}, \citenamefont {{Li}}, \citenamefont {{Miao}},\ and\ \citenamefont
  {{Li}}}]{2020SCPMA..6310412P}%
  \BibitemOpen
  \bibfield  {author} {\bibinfo {author} {\bibnamefont {{Pan}}, \bibfnamefont
  {S.}}, \bibinfo {author} {\bibfnamefont {M.}~\bibnamefont {{Liu}}}, \bibinfo
  {author} {\bibfnamefont {J.}~\bibnamefont {{Forero-Romero}}}, \bibinfo
  {author} {\bibfnamefont {C.~G.}\ \bibnamefont {{Sabiu}}}, \bibinfo {author}
  {\bibfnamefont {Z.}~\bibnamefont {{Li}}}, \bibinfo {author} {\bibfnamefont
  {H.}~\bibnamefont {{Miao}}}, and\ \bibinfo {author} {\bibfnamefont {X.-D.}\
  \bibnamefont {{Li}}}} (\bibinfo {year} {2020}),\ \href
  {https://doi.org/10.1007/s11433-020-1586-3} {\bibfield  {journal} {\bibinfo
  {journal} {Science China Physics, Mechanics, and Astronomy}\ }\textbf
  {\bibinfo {volume} {63}}~(\bibinfo {number} {11}),\ \bibinfo {eid}
  {110412}},\ \Eprint {https://arxiv.org/abs/1908.10590} {arXiv:1908.10590
  [astro-ph.CO]} \BibitemShut {NoStop}%
\bibitem [{\citenamefont {Pedregosa}\ \emph {et~al.}(2011)\citenamefont
  {Pedregosa}, \citenamefont {Varoquaux}, \citenamefont {Gramfort},
  \citenamefont {Michel}, \citenamefont {Thirion}, \citenamefont {Grisel},
  \citenamefont {Blondel}, \citenamefont {Prettenhofer}, \citenamefont {Weiss},
  \citenamefont {Dubourg} \emph {et~al.}}]{pedregosa2011scikit}%
  \BibitemOpen
  \bibfield  {author} {\bibinfo {author} {\bibnamefont {Pedregosa},
  \bibfnamefont {F.}}, \bibinfo {author} {\bibfnamefont {G.}~\bibnamefont
  {Varoquaux}}, \bibinfo {author} {\bibfnamefont {A.}~\bibnamefont {Gramfort}},
  \bibinfo {author} {\bibfnamefont {V.}~\bibnamefont {Michel}}, \bibinfo
  {author} {\bibfnamefont {B.}~\bibnamefont {Thirion}}, \bibinfo {author}
  {\bibfnamefont {O.}~\bibnamefont {Grisel}}, \bibinfo {author} {\bibfnamefont
  {M.}~\bibnamefont {Blondel}}, \bibinfo {author} {\bibfnamefont
  {P.}~\bibnamefont {Prettenhofer}}, \bibinfo {author} {\bibfnamefont
  {R.}~\bibnamefont {Weiss}}, \bibinfo {author} {\bibfnamefont
  {V.}~\bibnamefont {Dubourg}},  \emph {et~al.}} (\bibinfo {year} {2011}),\
  \href@noop {} {\bibfield  {journal} {\bibinfo  {journal} {the Journal of
  machine Learning research}\ }\textbf {\bibinfo {volume} {12}},\ \bibinfo
  {pages} {2825}}\BibitemShut {NoStop}%
\bibitem [{\citenamefont {Plumerault}\ \emph {et~al.}(2021)\citenamefont
  {Plumerault}, \citenamefont {Le~Borgne},\ and\ \citenamefont
  {Hudelot}}]{plumerault2021avae}%
  \BibitemOpen
  \bibfield  {author} {\bibinfo {author} {\bibnamefont {Plumerault},
  \bibfnamefont {A.}}, \bibinfo {author} {\bibfnamefont {H.}~\bibnamefont
  {Le~Borgne}}, and\ \bibinfo {author} {\bibfnamefont {C.}~\bibnamefont
  {Hudelot}}} (\bibinfo {year} {2021}),\ in\ \href@noop {} {\emph {\bibinfo
  {booktitle} {2020 25th International Conference on Pattern Recognition
  (ICPR)}}}\ (\bibinfo {organization} {IEEE})\ pp.\ \bibinfo {pages}
  {8687--8694}\BibitemShut {NoStop}%
\bibitem [{\citenamefont {Ramachandran}\ \emph {et~al.}(2017)\citenamefont
  {Ramachandran}, \citenamefont {Zoph},\ and\ \citenamefont
  {Le}}]{ramachandran2017searching}%
  \BibitemOpen
  \bibfield  {author} {\bibinfo {author} {\bibnamefont {Ramachandran},
  \bibfnamefont {P.}}, \bibinfo {author} {\bibfnamefont {B.}~\bibnamefont
  {Zoph}}, and\ \bibinfo {author} {\bibfnamefont {Q.~V.}\ \bibnamefont {Le}}}
  (\bibinfo {year} {2017}),\ \href@noop {} {\bibinfo  {journal} {arXiv preprint
  arXiv:1710.05941}\ }\BibitemShut {NoStop}%
\bibitem [{\citenamefont {Ratsimbazafy}\ \emph {et~al.}(2017)\citenamefont
  {Ratsimbazafy}, \citenamefont {Loubser}, \citenamefont {Crawford},
  \citenamefont {Cress}, \citenamefont {Bassett}, \citenamefont {Nichol},\ and\
  \citenamefont {V{\"a}is{\"a}nen}}]{ratsimbazafy2017age}%
  \BibitemOpen
\bibfield  {journal} {  }\bibfield  {author} {\bibinfo {author} {\bibnamefont
  {Ratsimbazafy}, \bibfnamefont {A.}}, \bibinfo {author} {\bibfnamefont
  {S.}~\bibnamefont {Loubser}}, \bibinfo {author} {\bibfnamefont
  {S.}~\bibnamefont {Crawford}}, \bibinfo {author} {\bibfnamefont
  {C.}~\bibnamefont {Cress}}, \bibinfo {author} {\bibfnamefont
  {B.}~\bibnamefont {Bassett}}, \bibinfo {author} {\bibfnamefont
  {R.}~\bibnamefont {Nichol}}, and\ \bibinfo {author} {\bibfnamefont
  {P.}~\bibnamefont {V{\"a}is{\"a}nen}}} (\bibinfo {year} {2017}),\ \href@noop
  {} {\bibfield  {journal} {\bibinfo  {journal} {Monthly Notices of the Royal
  Astronomical Society}\ }\textbf {\bibinfo {volume} {467}}~(\bibinfo {number}
  {3}),\ \bibinfo {pages} {3239}}\BibitemShut {NoStop}%
\bibitem [{\citenamefont {Schulz}\ \emph {et~al.}(2018)\citenamefont {Schulz},
  \citenamefont {Speekenbrink},\ and\ \citenamefont
  {Krause}}]{schulz2018tutorial}%
  \BibitemOpen
  \bibfield  {author} {\bibinfo {author} {\bibnamefont {Schulz}, \bibfnamefont
  {E.}}, \bibinfo {author} {\bibfnamefont {M.}~\bibnamefont {Speekenbrink}},
  and\ \bibinfo {author} {\bibfnamefont {A.}~\bibnamefont {Krause}}} (\bibinfo
  {year} {2018}),\ \href@noop {} {\bibfield  {journal} {\bibinfo  {journal}
  {Journal of mathematical psychology}\ }\textbf {\bibinfo {volume} {85}},\
  \bibinfo {pages} {1}}\BibitemShut {NoStop}%
\bibitem [{\citenamefont {Scolnic}\ \emph {et~al.}(2017)\citenamefont
  {Scolnic}, \citenamefont {Jones}, \citenamefont {Rest}, \citenamefont {Pan},
  \citenamefont {Chornock}, \citenamefont {Foley}, \citenamefont {Huber},
  \citenamefont {Kessler}, \citenamefont {Narayan},\ and\ \citenamefont
  {Riess}}]{2017The}%
  \BibitemOpen
  \bibfield  {author} {\bibinfo {author} {\bibnamefont {Scolnic}, \bibfnamefont
  {D.~M.}}, \bibinfo {author} {\bibfnamefont {D.~O.}\ \bibnamefont {Jones}},
  \bibinfo {author} {\bibfnamefont {A.}~\bibnamefont {Rest}}, \bibinfo {author}
  {\bibfnamefont {Y.~C.}\ \bibnamefont {Pan}}, \bibinfo {author} {\bibfnamefont
  {R.}~\bibnamefont {Chornock}}, \bibinfo {author} {\bibfnamefont {R.~J.}\
  \bibnamefont {Foley}}, \bibinfo {author} {\bibfnamefont {M.~E.}\ \bibnamefont
  {Huber}}, \bibinfo {author} {\bibfnamefont {R.}~\bibnamefont {Kessler}},
  \bibinfo {author} {\bibfnamefont {G.}~\bibnamefont {Narayan}}, and\ \bibinfo
  {author} {\bibfnamefont {A.~G.}\ \bibnamefont {Riess}}} (\bibinfo {year}
  {2017}),\ \href@noop {} {\bibinfo  {journal} {Astrophysical Journal}\
  }\BibitemShut {NoStop}%
\bibitem [{\citenamefont {Shyam}(2021)}]{shyam2021convolutional}%
  \BibitemOpen
\bibfield  {journal} {  }\bibfield  {author} {\bibinfo {author} {\bibnamefont
  {Shyam}, \bibfnamefont {R.}}} (\bibinfo {year} {2021}),\ \href@noop {}
  {\bibfield  {journal} {\bibinfo  {journal} {Journal of Computer Technology \&
  Applications}\ }\textbf {\bibinfo {volume} {12}}~(\bibinfo {number} {2}),\
  \bibinfo {pages} {6}}\BibitemShut {NoStop}%
\bibitem [{\citenamefont {Simon}\ \emph {et~al.}(2005)\citenamefont {Simon},
  \citenamefont {Verde},\ and\ \citenamefont {Jimenez}}]{PhysRevD.71.123001}%
  \BibitemOpen
  \bibfield  {author} {\bibinfo {author} {\bibnamefont {Simon}, \bibfnamefont
  {J.}}, \bibinfo {author} {\bibfnamefont {L.}~\bibnamefont {Verde}}, and\
  \bibinfo {author} {\bibfnamefont {R.}~\bibnamefont {Jimenez}}} (\bibinfo
  {year} {2005}),\ \href {https://doi.org/10.1103/PhysRevD.71.123001}
  {\bibfield  {journal} {\bibinfo  {journal} {Phys. Rev. D}\ }\textbf {\bibinfo
  {volume} {71}},\ \bibinfo {pages} {123001}}\BibitemShut {NoStop}%
\bibitem [{\citenamefont {Stern}\ \emph {et~al.}(2009)\citenamefont {Stern},
  \citenamefont {Jimenez}, \citenamefont {Verde}, \citenamefont
  {Kamionkowski},\ and\ \citenamefont {Stanford}}]{2009Cosmic}%
  \BibitemOpen
  \bibfield  {author} {\bibinfo {author} {\bibnamefont {Stern}, \bibfnamefont
  {D.}}, \bibinfo {author} {\bibfnamefont {R.}~\bibnamefont {Jimenez}},
  \bibinfo {author} {\bibfnamefont {L.}~\bibnamefont {Verde}}, \bibinfo
  {author} {\bibfnamefont {M.}~\bibnamefont {Kamionkowski}}, and\ \bibinfo
  {author} {\bibfnamefont {S.~A.}\ \bibnamefont {Stanford}}} (\bibinfo {year}
  {2009}),\ \href@noop {} {\bibinfo  {journal} {journal of cosmology and
  astroparticle physics}\ }\BibitemShut {NoStop}%
\bibitem [{\citenamefont {Tao}\ \emph {et~al.}(2022)\citenamefont {Tao},
  \citenamefont {Zhao}, \citenamefont {Zhang}, \citenamefont {Gajjar},
  \citenamefont {Zhu}, \citenamefont {Yue}, \citenamefont {Zhang},
  \citenamefont {Liu}, \citenamefont {Li}, \citenamefont {Zhang} \emph
  {et~al.}}]{tao2022sensitive}%
  \BibitemOpen
\bibfield  {journal} {  }\bibfield  {author} {\bibinfo {author} {\bibnamefont
  {Tao}, \bibfnamefont {Z.-Z.}}, \bibinfo {author} {\bibfnamefont {H.-C.}\
  \bibnamefont {Zhao}}, \bibinfo {author} {\bibfnamefont {T.-J.}\ \bibnamefont
  {Zhang}}, \bibinfo {author} {\bibfnamefont {V.}~\bibnamefont {Gajjar}},
  \bibinfo {author} {\bibfnamefont {Y.}~\bibnamefont {Zhu}}, \bibinfo {author}
  {\bibfnamefont {Y.-L.}\ \bibnamefont {Yue}}, \bibinfo {author} {\bibfnamefont
  {H.-Y.}\ \bibnamefont {Zhang}}, \bibinfo {author} {\bibfnamefont {W.-F.}\
  \bibnamefont {Liu}}, \bibinfo {author} {\bibfnamefont {S.-Y.}\ \bibnamefont
  {Li}}, \bibinfo {author} {\bibfnamefont {J.-C.}\ \bibnamefont {Zhang}},
  \emph {et~al.}} (\bibinfo {year} {2022}),\ \href@noop {} {\bibfield
  {journal} {\bibinfo  {journal} {The Astronomical Journal}\ }\textbf {\bibinfo
  {volume} {164}}~(\bibinfo {number} {4}),\ \bibinfo {pages} {160}}\BibitemShut
  {NoStop}%
\bibitem [{\citenamefont {Tosi}(2009)}]{tosi2009matplotlib}%
  \BibitemOpen
  \bibfield  {author} {\bibinfo {author} {\bibnamefont {Tosi}, \bibfnamefont
  {S.}}} (\bibinfo {year} {2009}),\ \href@noop {} {\emph {\bibinfo {title}
  {Matplotlib for Python developers}}}\ (\bibinfo  {publisher} {Packt
  Publishing Ltd})\BibitemShut {NoStop}%
\bibitem [{\citenamefont {Virtanen}\ \emph {et~al.}(2020)\citenamefont
  {Virtanen}, \citenamefont {Gommers}, \citenamefont {Oliphant}, \citenamefont
  {Haberland}, \citenamefont {Reddy}, \citenamefont {Cournapeau}, \citenamefont
  {Burovski}, \citenamefont {Peterson}, \citenamefont {Weckesser},
  \citenamefont {Bright} \emph {et~al.}}]{virtanen2020scipy}%
  \BibitemOpen
  \bibfield  {author} {\bibinfo {author} {\bibnamefont {Virtanen},
  \bibfnamefont {P.}}, \bibinfo {author} {\bibfnamefont {R.}~\bibnamefont
  {Gommers}}, \bibinfo {author} {\bibfnamefont {T.~E.}\ \bibnamefont
  {Oliphant}}, \bibinfo {author} {\bibfnamefont {M.}~\bibnamefont {Haberland}},
  \bibinfo {author} {\bibfnamefont {T.}~\bibnamefont {Reddy}}, \bibinfo
  {author} {\bibfnamefont {D.}~\bibnamefont {Cournapeau}}, \bibinfo {author}
  {\bibfnamefont {E.}~\bibnamefont {Burovski}}, \bibinfo {author}
  {\bibfnamefont {P.}~\bibnamefont {Peterson}}, \bibinfo {author}
  {\bibfnamefont {W.}~\bibnamefont {Weckesser}}, \bibinfo {author}
  {\bibfnamefont {J.}~\bibnamefont {Bright}},  \emph {et~al.}} (\bibinfo {year}
  {2020}),\ \href@noop {} {\bibfield  {journal} {\bibinfo  {journal} {Nature
  methods}\ }\textbf {\bibinfo {volume} {17}}~(\bibinfo {number} {3}),\
  \bibinfo {pages} {261}}\BibitemShut {NoStop}%
\bibitem [{\citenamefont {Wang}\ \emph {et~al.}(2020)\citenamefont {Wang},
  \citenamefont {Ma}, \citenamefont {Li},\ and\ \citenamefont
  {Xia}}]{wang2020reconstructing}%
  \BibitemOpen
  \bibfield  {author} {\bibinfo {author} {\bibnamefont {Wang}, \bibfnamefont
  {G.-J.}}, \bibinfo {author} {\bibfnamefont {X.-J.}\ \bibnamefont {Ma}},
  \bibinfo {author} {\bibfnamefont {S.-Y.}\ \bibnamefont {Li}}, and\ \bibinfo
  {author} {\bibfnamefont {J.-Q.}\ \bibnamefont {Xia}}} (\bibinfo {year}
  {2020}),\ \href@noop {} {\bibfield  {journal} {\bibinfo  {journal} {The
  Astrophysical Journal Supplement Series}\ }\textbf {\bibinfo {volume}
  {246}}~(\bibinfo {number} {1}),\ \bibinfo {pages} {13}}\BibitemShut {NoStop}%
\bibitem [{\citenamefont {Wang}\ and\ \citenamefont
  {Zhang}(2011)}]{2011Constraints}%
  \BibitemOpen
  \bibfield  {author} {\bibinfo {author} {\bibnamefont {Wang}, \bibfnamefont
  {H.}}, and\ \bibinfo {author} {\bibfnamefont {T.~J.}\ \bibnamefont {Zhang}}}
  (\bibinfo {year} {2011}),\ \href@noop {} {\bibfield  {journal} {\bibinfo
  {journal} {The Astrophysical Journal}\ }\textbf {\bibinfo {volume}
  {748}}~(\bibinfo {number} {2}),\ \bibinfo {pages} {315}}\BibitemShut
  {NoStop}%
\bibitem [{\citenamefont {Wang}\ \emph {et~al.}(2021)\citenamefont {Wang},
  \citenamefont {Xie}, \citenamefont {Zhang}, \citenamefont {Huang},
  \citenamefont {Zhang},\ and\ \citenamefont {Liu}}]{2021Likelihood}%
  \BibitemOpen
  \bibfield  {author} {\bibinfo {author} {\bibnamefont {Wang}, \bibfnamefont
  {Y.~C.}}, \bibinfo {author} {\bibfnamefont {Y.~B.}\ \bibnamefont {Xie}},
  \bibinfo {author} {\bibfnamefont {T.~J.}\ \bibnamefont {Zhang}}, \bibinfo
  {author} {\bibfnamefont {H.~C.}\ \bibnamefont {Huang}}, \bibinfo {author}
  {\bibfnamefont {T.}~\bibnamefont {Zhang}}, and\ \bibinfo {author}
  {\bibfnamefont {K.}~\bibnamefont {Liu}}} (\bibinfo {year} {2021}),\
  \href@noop {} {\bibfield  {journal} {\bibinfo  {journal} {The Astrophysical
  Journal Supplement Series}\ }\textbf {\bibinfo {volume} {254}}~(\bibinfo
  {number} {2}),\ \bibinfo {pages} {43 (16pp)}}\BibitemShut {NoStop}%
\bibitem [{\citenamefont {Williams}\ and\ \citenamefont
  {Rasmussen}(1995)}]{williams1995gaussian}%
  \BibitemOpen
  \bibfield  {author} {\bibinfo {author} {\bibnamefont {Williams},
  \bibfnamefont {C.}}, and\ \bibinfo {author} {\bibfnamefont {C.}~\bibnamefont
  {Rasmussen}}} (\bibinfo {year} {1995}),\ \href@noop {} {\bibfield  {journal}
  {\bibinfo  {journal} {Advances in neural information processing systems}\
  }\textbf {\bibinfo {volume} {8}}}\BibitemShut {NoStop}%
\bibitem [{\citenamefont {Yamashita}\ \emph {et~al.}(2018)\citenamefont
  {Yamashita}, \citenamefont {Nishio}, \citenamefont {Do},\ and\ \citenamefont
  {Togashi}}]{yamashita2018convolutional}%
  \BibitemOpen
  \bibfield  {author} {\bibinfo {author} {\bibnamefont {Yamashita},
  \bibfnamefont {R.}}, \bibinfo {author} {\bibfnamefont {M.}~\bibnamefont
  {Nishio}}, \bibinfo {author} {\bibfnamefont {R.~K.~G.}\ \bibnamefont {Do}},
  and\ \bibinfo {author} {\bibfnamefont {K.}~\bibnamefont {Togashi}}} (\bibinfo
  {year} {2018}),\ \href@noop {} {\bibfield  {journal} {\bibinfo  {journal}
  {Insights into imaging}\ }\textbf {\bibinfo {volume} {9}},\ \bibinfo {pages}
  {611}}\BibitemShut {NoStop}%
\bibitem [{\citenamefont {Yegnanarayana}(2009)}]{yegnanarayana2009artificial}%
  \BibitemOpen
  \bibfield  {author} {\bibinfo {author} {\bibnamefont {Yegnanarayana},
  \bibfnamefont {B.}}} (\bibinfo {year} {2009}),\ \href@noop {} {\emph
  {\bibinfo {title} {Artificial neural networks}}}\ (\bibinfo  {publisher} {PHI
  Learning Pvt. Ltd.})\BibitemShut {NoStop}%
\bibitem [{\citenamefont {You}\ \emph {et~al.}(2017)\citenamefont {You},
  \citenamefont {Gitman},\ and\ \citenamefont {Ginsburg}}]{you2017large}%
  \BibitemOpen
  \bibfield  {author} {\bibinfo {author} {\bibnamefont {You}, \bibfnamefont
  {Y.}}, \bibinfo {author} {\bibfnamefont {I.}~\bibnamefont {Gitman}}, and\
  \bibinfo {author} {\bibfnamefont {B.}~\bibnamefont {Ginsburg}}} (\bibinfo
  {year} {2017}),\ \href@noop {} {\bibinfo  {journal} {arXiv preprint
  arXiv:1708.03888}\ }\BibitemShut {NoStop}%
\bibitem [{\citenamefont {Yu}\ \emph {et~al.}(2017)\citenamefont {Yu},
  \citenamefont {Emberson}, \citenamefont {Inman}, \citenamefont {Zhang},
  \citenamefont {Pen}, \citenamefont {Harnois-D{\'e}raps}, \citenamefont
  {Yuan}, \citenamefont {Teng}, \citenamefont {Zhu}, \citenamefont {Chen} \emph
  {et~al.}}]{yu2017differential}%
  \BibitemOpen
\bibfield  {journal} {  }\bibfield  {author} {\bibinfo {author} {\bibnamefont
  {Yu}, \bibfnamefont {H.-R.}}, \bibinfo {author} {\bibfnamefont
  {J.}~\bibnamefont {Emberson}}, \bibinfo {author} {\bibfnamefont
  {D.}~\bibnamefont {Inman}}, \bibinfo {author} {\bibfnamefont {T.-J.}\
  \bibnamefont {Zhang}}, \bibinfo {author} {\bibfnamefont {U.-L.}\ \bibnamefont
  {Pen}}, \bibinfo {author} {\bibfnamefont {J.}~\bibnamefont
  {Harnois-D{\'e}raps}}, \bibinfo {author} {\bibfnamefont {S.}~\bibnamefont
  {Yuan}}, \bibinfo {author} {\bibfnamefont {H.-Y.}\ \bibnamefont {Teng}},
  \bibinfo {author} {\bibfnamefont {H.-M.}\ \bibnamefont {Zhu}}, \bibinfo
  {author} {\bibfnamefont {X.}~\bibnamefont {Chen}},  \emph {et~al.}} (\bibinfo
  {year} {2017}),\ \href@noop {} {\bibfield  {journal} {\bibinfo  {journal}
  {Nature Astronomy}\ }\textbf {\bibinfo {volume} {1}}~(\bibinfo {number}
  {7}),\ \bibinfo {pages} {0143}}\BibitemShut {NoStop}%
\bibitem [{\citenamefont {Yuan}\ and\ \citenamefont
  {Zhang}(2015)}]{yuan2015breaking}%
  \BibitemOpen
  \bibfield  {author} {\bibinfo {author} {\bibnamefont {Yuan}, \bibfnamefont
  {S.}}, and\ \bibinfo {author} {\bibfnamefont {T.-J.}\ \bibnamefont {Zhang}}}
  (\bibinfo {year} {2015}),\ \href@noop {} {\bibfield  {journal} {\bibinfo
  {journal} {Journal of Cosmology and Astroparticle Physics}\ }\textbf
  {\bibinfo {volume} {2015}}~(\bibinfo {number} {02}),\ \bibinfo {pages}
  {025}}\BibitemShut {NoStop}%
\bibitem [{\citenamefont {Zhang}\ \emph
  {et~al.}(2023{\natexlab{a}})\citenamefont {Zhang}, \citenamefont {Wang},
  \citenamefont {Zhang},\ and\ \citenamefont {Zhang}}]{zhang2023kernel}%
  \BibitemOpen
  \bibfield  {author} {\bibinfo {author} {\bibnamefont {Zhang}, \bibfnamefont
  {H.}}, \bibinfo {author} {\bibfnamefont {Y.-C.}\ \bibnamefont {Wang}},
  \bibinfo {author} {\bibfnamefont {T.-J.}\ \bibnamefont {Zhang}}, and\
  \bibinfo {author} {\bibfnamefont {T.}~\bibnamefont {Zhang}}} (\bibinfo {year}
  {2023}{\natexlab{a}}),\ \href@noop {} {\bibfield  {journal} {\bibinfo
  {journal} {The Astrophysical Journal Supplement Series}\ }\textbf {\bibinfo
  {volume} {266}}~(\bibinfo {number} {2}),\ \bibinfo {pages} {27}}\BibitemShut
  {NoStop}%
\bibitem [{\citenamefont {Zhang}\ \emph
  {et~al.}(2023{\natexlab{b}})\citenamefont {Zhang}, \citenamefont {Hu},
  \citenamefont {Jiao}, \citenamefont {Wang}, \citenamefont {Xie},
  \citenamefont {Yu}, \citenamefont {Zhao},\ and\ \citenamefont
  {Zhang}}]{zhang2023non}%
  \BibitemOpen
  \bibfield  {author} {\bibinfo {author} {\bibnamefont {Zhang}, \bibfnamefont
  {J.-C.}}, \bibinfo {author} {\bibfnamefont {Y.}~\bibnamefont {Hu}}, \bibinfo
  {author} {\bibfnamefont {K.}~\bibnamefont {Jiao}}, \bibinfo {author}
  {\bibfnamefont {H.-F.}\ \bibnamefont {Wang}}, \bibinfo {author}
  {\bibfnamefont {Y.-B.}\ \bibnamefont {Xie}}, \bibinfo {author} {\bibfnamefont
  {B.}~\bibnamefont {Yu}}, \bibinfo {author} {\bibfnamefont {L.-L.}\
  \bibnamefont {Zhao}}, and\ \bibinfo {author} {\bibfnamefont {T.-J.}\
  \bibnamefont {Zhang}}} (\bibinfo {year} {2023}{\natexlab{b}}),\ \href@noop {}
  {\bibinfo  {journal} {arXiv preprint arXiv:2311.13938}\ }\BibitemShut
  {NoStop}%
\bibitem [{\citenamefont {Zhang}\ \emph {et~al.}(2024)\citenamefont {Zhang},
  \citenamefont {Hu}, \citenamefont {Jiao}, \citenamefont {Wang}, \citenamefont
  {Xie}, \citenamefont {Yu}, \citenamefont {Zhao},\ and\ \citenamefont
  {Zhang}}]{Zhang_2024}%
  \BibitemOpen
\bibfield  {journal} {  }\bibfield  {author} {\bibinfo {author} {\bibnamefont
  {Zhang}, \bibfnamefont {J.-C.}}, \bibinfo {author} {\bibfnamefont
  {Y.}~\bibnamefont {Hu}}, \bibinfo {author} {\bibfnamefont {K.}~\bibnamefont
  {Jiao}}, \bibinfo {author} {\bibfnamefont {H.-F.}\ \bibnamefont {Wang}},
  \bibinfo {author} {\bibfnamefont {Y.-B.}\ \bibnamefont {Xie}}, \bibinfo
  {author} {\bibfnamefont {B.}~\bibnamefont {Yu}}, \bibinfo {author}
  {\bibfnamefont {L.-L.}\ \bibnamefont {Zhao}}, and\ \bibinfo {author}
  {\bibfnamefont {T.-J.}\ \bibnamefont {Zhang}}} (\bibinfo {year} {2024}),\
  \href {https://doi.org/10.3847/1538-4365/ad0f1e} {\bibfield  {journal}
  {\bibinfo  {journal} {The Astrophysical Journal Supplement Series}\ }\textbf
  {\bibinfo {volume} {270}}~(\bibinfo {number} {2}),\ \bibinfo {pages}
  {23}}\BibitemShut {NoStop}%
\end{thebibliography}%

\end{document}